\documentclass[preprint,showpacs,preprintnumbers,amsmath,amssymb]{revtex4}

\usepackage{amsmath}
\usepackage{amssymb}
\usepackage{amsthm}
\usepackage{amsfonts}
\usepackage{graphicx}
\usepackage{dcolumn}
\usepackage{bm}
\usepackage{epsfig}
\usepackage{mathrsfs}

\begin{document}
\title{Thermodynamics, spin-charge separation and correlation functions
  of spin-$1/2$ fermions with repulsive interaction}

\author{J. Y. Lee$^{1}$,
 X. W. Guan$^{1}$, K. Sakai$^{2}$
 and M. T. Batchelor$^{1, 3}$}

 \affiliation{$^{1}$ Department of Theoretical Physics,
 Research School of Physics and Engineering,
 Australian National University, Canberra ACT 0200, Australia}

 \affiliation{$^{2}$ Institute of Physics, University of Tokyo, Komaba 3-8-1,
 Meguro-ku, Tokyo 153-8902, Japan}

 \affiliation{$^{3}$ Mathematical Sciences Institute,
 Australian National University, Canberra ACT 0200, Australia}

\date{\today}

\begin{abstract}
We investigate the low temperature thermodynamics and correlation
functions of one-dimensional spin-1/2 fermions with strong repulsion
in an external magnetic field via the thermodynamic Bethe ansatz
method. The exact thermodynamics of the model in a weak magnetic
field is derived with the help of Wiener-Hopf techniques. It turns
out that the low energy physics can be described by spin-charge
separated conformal field theories of an effective
Tomonaga-Luttinger liquid and an antiferromagnetic $SU(2)$
Heisenberg spin chain. However, these two types of conformally
invariant low-lying excitations may break down as excitations take
place far away from the Fermi points. The long distance asymptotics
of the correlation functions and the critical exponents for the
model in the presence of a magnetic field at zero temperature are
derived in detail by solving dressed charge equations and by
conformal mapping. Furthermore, we calculate the conformal
dimensions for particular cases of correlation functions. The
leading terms of these correlation functions are given explicitly
for a weak magnetic field $H\ll 1$ and for a magnetic field close to
the critical field $H\rightarrow H_{c}$. Our analytical results
provide insights into universal thermodynamics and criticality in
one-dimensional many-body physics.

\end{abstract}



\maketitle

\section{Introduction}

Since the pioneering work in the 60's, 70's and 80's by  McGuire,
Yang, Lieb, Sutherland, Baxter \emph{et al.}, and of the St
Petersburg and Kyoto schools, the study of integrable models has
flourished into a major activity. Almost without exception, the
energy levels are given exactly in terms of the Bethe ansatz (BA)
equations, from which physical properties can be calculated. This is
a hallmark of integrable models that exhibit Yang-Baxter symmetry
\cite{Bethe1931}. The knowledge and understanding gained from  integrable
models have greatly enhanced progress in the theory of phase
transitions and critical phenomena. The most significant results
achieved to date have been for two-dimensional  lattice models
and their related one-dimensional (1D) quantum spin chains as well
as strongly correlated electronic systems
\cite{book1,tak,book2,book3,book4}. Integrable models
are also known for systems such as Bose-Einstein condensates
\cite{ZJMG,Cao}, metallic nano\-grains \cite{Siera} and impurity
models \cite{Konik,Mehta2006}.

In general, the BA solution for 1D integrable systems is a set of
coupled algebraic equations. Finding a set of solutions
of quasimomenta and spin rapidities $\{k_{j},\Lambda_{j}\}$ for the
BA equations gives the energy $\sum_{j}k_{j}^{2}$ and the momentum
$\sum_{j}k_{j}$ of the system. However, the BA equations by
themselves do not explicitly exhibit any temperature dependence. At
zero temperature, the BA equations in principle give the complete
eigenstates of the model. However, at finite temperatures, the
equilibrium states become degenerate. Thus the thermodynamics of
these BA solvable models are instead determined by a set of coupled
nonlinear integral equations called the thermodynamic Bethe ansatz
(TBA) equations \cite{Yang1969}. The TBA equations are expressed in
terms of the dressed energies of different ``Fermi seas'' that are
functions of temperature, chemical potential and external magnetic
fields. The TBA equations in the zero temperature limit, i.e., $T\to
0$, give rise to the so called dressed energy equations which
describe the band fillings with respect to Zeeman fields and
chemical potentials.

The TBA equations are very
difficult to solve in general. They involve an infinite number of
coupled nonlinear integral equations for spin strings which are
quite cumbersome to solve using either analytical or
numerical methods \cite{tak,book3}.  Recently,  Caux  {\em et al}  \cite{Caux1,Caux2} developed numerical  schemes to solve the TBA equations of the 1D two-component spinor  Bose gas with delta-function interaction.  The results obtained by these numerical schemes show an insightful   interplay between  quantum statistics, interactions and temperature in 1D interacting many-body systems.  In the context of the quantum transfer matrix method \cite{QTM}, Kl\"{u}mper  and Patu  \cite{Klumper}  derived the nonlinear integral equations for the 1D Bose and Fermi gases with repulsive delta-function interaction.  This approach  opens up the possibility of obtaining the thermodynamics of the continuum models of  interacting fermions and bosons by taking an appropriate limit for the integrable lattice models. The advance of such  approaches is the reduction of the infinite number of  TBA equations to a  finite number of the nonlinear integral equations. Where the finite number of the nonlinear integral equations for the lattice models can be solved  numerically and analytically in certain temperature regimes \cite{BGOT,T-system}.   Despite giving  high precision  numerical thermodynamics, finding the universal nature of interacting particles requires  further analytical input. Significant universal features  of 1D many-body systems are   Tomonaga-Luttinger liquid physics and quantum critical phenomena at low temperatures which  involve finding essentially   universal parameters such as central charges, Luttinger parameters, correlation exponents and  dynamical critical exponents.   All  studies of these universal parameters  call for  mathematical  analysis and analytical derivation.  

Some progress
has been made to derive low temperature analytic results for BA
solvable models.  For example, Mezincescu \emph{et al.}
\cite{Nepomechie1992,Nepomechie1993} obtained the free energy of
spin chains at low temperatures  under a small magnetic field by
using the Wiener-Hopf technique. Johnson and McCoy
\cite{Johnson1972} obtained the leading  temperature dependent terms
in a low temperature expansion of the free energy for the massive
regime of the Heisenberg model. Filyov \emph{et al.}
\cite{Filyov1981} gave an exact solution to the s-d exchange model
expressed as a series in terms of the temperature. Some analytical results for the
TBA equations of 1D many-body systems are restricted to the ground
state ($T=0$) in the strong coupling limit ($c\gg 1$)
\cite{Penc,GB,Wadati}. 
Recently, further
progress has been made to obtain the analytic finite temperature
thermodynamics and quantum criticality  of 1D attractive fermions with
strongly attractive interactions \cite{Zhao2009,Guan2010,GHo,GBat}.

(1+1)-dimensional critical systems  
not only have global scale invariance but exhibit local scale
invariance (conformal invariance) too. The conformal group in
(1+1)-dimensions is infinite dimensional and completely determines
the critical exponents and bulk correlation functions at criticality
for gapless excitations \cite{Belavin1984}. Close to criticality,
the dispersion relations for 1D quantum systems are approximately
linear. Conformal invariance predicts that the energy per unit
length has a universal finite size scaling form $E=E_{0}+\Delta/L^2$
where $E_{0}$ is the ground state energy per unit length for the
infinite system and $\Delta$ is a universal term. These universality
classes are characterized by the dimensionless number $C$ (contained
in the term $\Delta$) which is the central charge of the underlying
Virasoro algebra \cite{Affleck1986,Blote1986}. Affleck
\cite{Affleck1986} also showed that conformal invariance gives a
universal form for the finite temperature effects on the free energy
by replacing $1/L$ with $T$ in the conformal map
$z=\exp(2\pi\omega/L)$. At the same time, Cardy \cite{Cardy1986}
showed that the two-point correlation function between primary
fields can be directly derived from conformal mapping using transfer
matrix techniques and expressed the conformal dimensions in terms of
finite-size corrections to the energy spectrum. When $C<1$, it takes
on discrete values only i.e., $C$ is quantized and hence the
conformal dimensions are restricted to certain rational numbers
\cite{Friedan1984}. On the other hand when $C\geq 1$, the critical
exponents may depend continuously on the parameters of the model
\cite{Zamolodchikov1985}.

The critical exponents for BA integrable models can be calculated
via the quantum inverse scattering method (QISM) in terms of
a function $Z(\lambda)$ called the dressed charge. In this way
Bogoliubov \emph{et al.} \cite{Bogoliubov1986} obtained
explicit expressions for the correlation functions for the Bose gas and the
XXX and XXZ chains. Izergin \emph{et al.}
\cite{Izergin1989} considered the finite-size corrections to
multicomponent BA systems and presented a formula for the dressed
charge matrix $Z_{\alpha\beta}$ which determines the critical
exponents. They also showed that the integral equations for the
dressed charge matrix depend only on the quantum $R$-matrix of the
model, which means that the universal class of critical exponents
are described by the $R$-matrix and the structure of the ground
states of the integrable models. This universality property is a
consequence of conformal invariance. Other models like the
impenetrable Bose gas \cite{Its1990}, the supersymmetric $t-J$ model
\cite{Kawakami1991} and the Hubbard model
\cite{Frahm1990,Frahm1991,Woynarovich1987,Woynarovich1989} have also
been considered in the context of the QISM approach. The Luttinger
liquid  is an alternative approach based on the fact that these
models are certain realizations of the Gaussian model
\cite{Haldane1981a,Haldane1981b}. Progress has also been made using
Fredholm determinant representations of time-dependent temperature
correlation functions for bosons and fermions in 1D when $c=\infty$
\cite{Izergin1997,Izergin1998}.

In this paper, we focus on the universal nature of 1D repulsive spin-1/2 fermions  in the frame work of 
the TBA formalism, including Luttinger physics and critical behaviour of correlation functions.  In order to elucidate  the significant features of spin-charge separation and critical  exponents at quantum criticality, it is essential to analytically calculate the dressed energy potentials which encode the quantum and thermal  fluctuations of the spin and charge degrees of freedom in the critical regime.  We investigate the low temperature thermodynamics of
strongly repulsive spin-1/2 fermions in a small magnetic field and a magnetic field close to the saturation field via
the TBA method. We take an approximation to the TBA equations in the
strong coupling regime, where the interacting strength $c\gg 1$.
Thus the TBA equations are transformed into a new set of equations
which can be solved using the Wiener-Hopf method. A comparison of
the pressure and the entropy is made between the application of two
different integral expansions. These are Sommerfeld's lemma, which is
valid for very low temperatures, and the polylogarithm function which
is valid for finite temperatures. The result from Sommerfeld
expansion agrees with the conformal field theory prediction
\cite{Affleck1986,Blote1986}. It is shown that the low energy
physics can be described by a spin-charge separated theory of an
effective Tomonaga-Luttinger liquid and antiferromagnetic $SU(2)$
Heisenberg spin chain. A universal crossover from a relativistic
dispersion to a nonrelativistic dispersion is determined by the
exact thermodynamics extracted from the polylogarithm function. We
also derive the explicit dressed charge matrix elements for the
model in a weak external field $H\ll 1$ using the Wiener-Hopf method
again, and also for the case that is close to the ferromagnetic
state $H\rightarrow H_{c}$, where $H_{c}$ is the critical magnetic
field. Various two-point and multi-point correlation functions at
zero temperature are derived based on the expressions obtained from
the dressed charge matrix. The leading terms and their critical
exponents are given explicitly. Our results show that there is
indeed no long-range order in this system.

The spin-1/2 fermion model under consideration is the 
continuum limit of the 1D Hubbard model (see, e.g., pp 45-49 of Ref.~\cite{book3}), 
which has been widely studied \cite{book3}. In particular, the various correlation functions 
and scaling dimensions obtained here for the spin-1/2 fermion model have been 
derived for the 1D Hubbard model, for interacting fermions and for a mixture of bosons and fermions    using the
dressed charge formalism \cite{book3,Frahm1990,Frahm1991,Frahm-P,Lee-Guan}. Accordingly our results 
in the infinite coupling limit reduce to those  obtained for the 1D Hubbard model with an infinitely strong repulsion.  
Caution should be paid to the order of the limits $T\to 0$ and infinite strong coupling  \cite{Cheianov}.
Taking the $T\to 0$ limit first,  the correlations (for example, the one-particle correlation)    
 show the scaling behavior of conformal field theory in the infinite strong coupling limit. However, taking the infinite strong coupling limit first, 
 the correlations decay exponentially in the $T\to 0$ limit. The two limits do not commute. 
Moreover, in the grand canonical ensemble, the Tomonaga-Luttinger liquid exists only  in a certain region where the chemical potential is greater 
than the critical  value. Below the critical chemical potential, the low temperature thermodynamics is that of an 
ideal gas in another regime, see \cite{Gohmann}.

The corrections we have obtained in terms of strong but finite coupling 
are of importance because of the experimental developments which enable 
access to the finitely strong coupling regime \cite{nature}.  
Our analytical $1/c$ order corrections to the critical exponents indicate an important signature 
--  the critical exponents depend on the model parameters with  central charge $C \ge 1$.
In addition, our thermodynamical properties are valid for temperatures from $T=0$ to  $T \ll  c^2$. 
Results of this kind are necessary to 
test critical phenomena and spin-charge separation theory in experiments with 
trapped fermionic atoms. 
Indeed, along with extending the known results for the thermodynamics and correlations, 
this is our main motivation here.

This paper is set out as follows. In Section \ref{section-TBA} we
introduce the model and present the TBA equations. The low
temperature thermodynamics is derived in Section \ref{therm} by
expressing the TBA equations in the form of Wiener-Hopf integral
equations. We solve the dressed charge equations for the model in
the small field limit ($H\ll 1$) and in the limit where the field
approaches the critical value ($H\rightarrow H_{c}$) in Section
\ref{section-dressed}. In Section \ref{section-corr}, we calculate the
correlation functions of various operators in both limits. We then
give a summary of our main results and concluding remarks in Section
\ref{section-conclude}.  Some detailed working and results are given in the  Appendices. In Appendix  \ref{ground}, we derive the
ground state thermodynamics and the critical field for the model.
The Wiener-Hopf method is discussed in Appendix \ref{WH}.  A more detailed derivation of the low temperature
thermodynamics is given in Appendix \ref{low-temp}. The leading terms of the zero temperature correlation functions are given in Appendices \ref{CF-1} and \ref{CF-1}. 

\section{The TBA equations}\label{section-TBA}

We consider a system of 1D spin-1/2 fermions with delta-function
interaction, with hamiltonian
\begin{equation}
\mathscr{H}=\mathscr{H}_0-\mu N, \qquad \mathscr{H}_0=-\sum_{j=1}^N
\frac{\partial}{\partial x_j^2}+ 2c \sum_{1\le j<k\le
N}\delta(x_j-x_k)-H M. \label{hamiltonian}
\end{equation}
Here $N=N_{\uparrow}+N_{\downarrow}$ is the total number of spin-up $N_{\uparrow}$ and
spin-down $N_{\downarrow}$ fermions and
$M=(N_{\uparrow}-N_{\downarrow})/2$ is the magnetization.
 $\mu$ and $H$ are the chemical potential and the magnetic field.
 In this paper, we
exclusively consider the case of repulsive interaction for which $c>0$.

The ground state properties can be obtained using the BA
solution \cite{Yang,Gaudin} (see Appendix~\ref{ground} for a brief
review). On the other hand, at finite temperatures $T>0$, the
physical quantities are described by the following set of non-linear
integral equations, which are referred to as the TBA equations
\cite{tak,Lai}. In the thermodynamic limit, their explicit form is
\begin{align}
\varepsilon(k)&=k^{2}-\mu-\frac{H}{2}-T\sum_{n=1}^{\infty}a_{n}\ast\ln\left(1+e^{-\phi_{n}(k)/T}\right)\label{TBA-k}\\
\phi_{n}(\lambda)&=nH-Ta_{n}\ast\ln\left(1+e^{-\varepsilon(\lambda)/T}\right)
+T\sum_{m=1}^{\infty}T_{nm}\ast\ln\left(1+e^{-\phi_{m}(\lambda)/T}\right)
\label{TBA-lambda}
\end{align}
or equivalently
\begin{align}
\varepsilon(k)&=k^{2}-\mu-T K
\ast\ln(1+e^{-\varepsilon(k)/T})-Ts\ast\ln\left(1+e^{\phi_{1}(k)/T}\right)
\label{TBA-epsilon-o}\\
\phi_{1}(\lambda)&=Ts\ast\ln\left(1+e^{\phi_{2}(\lambda)/T}\right)
-Ts\ast\ln\left(1+e^{-\varepsilon(\lambda)/T}\right)\label{TBA-phi-o}\\
\phi_{n}(\lambda)&=Ts\ast\ln\left(1+e^{\phi_{n-1}(\lambda)/T}\right)+
Ts\ast\ln\left(1+e^{\phi_{n+1}(\lambda)/T}\right) \label{TBA-phi-n}
\end{align}
where the functions $\phi_n(\lambda)$ must satisfy the condition
\begin{equation}
\lim_{n\rightarrow\infty}\frac{\phi_{n}(\lambda)}{n}=H.
\end{equation}

The functions $a=a_{n}(\lambda)$, $s=s(\lambda)$, $K=K(\lambda)$ and
$T_{nm} = T_{nm}(\lambda)$ are defined by
\begin{align}
&a_{n}(\lambda)=\frac{1}{\pi}\frac{nc/2}{(nc)^{2}/4+\lambda^{2}},
\quad s(\lambda)=\frac{1}{2c\cosh(\pi \lambda/c)}, \quad
K(\lambda)=\frac{1}{2\pi}\int_{-\infty}^{\infty}
\frac{1}{1+e^{c|\omega|}}
e^{-i \omega \lambda} d \omega,
  \nonumber \\
&T_{nm}(\lambda)=\begin{cases}
              a_{|n-m|}(\lambda)+2a_{|n-m|+2}(\lambda)+\ldots+2a_{n+m-2}(\lambda)+a_{n+m}(\lambda) &
                                                     \text{ for $n\neq m$}; \\
              2a_{2}(\lambda)+2a_{4}(\lambda)+\ldots+2a_{2n-2}(\lambda)+a_{2n}(\lambda) & \text{ for
              $n=m$}.
            \end{cases}
\label{kernel}
\end{align}
The asterisk $*$ denotes the convolution
$f\ast g(x)=\int_{-\infty}^{\infty}f(x-x')g(x')dx'$.

The bulk quantities are characterized by the solution to the TBA
equations. For instance, the  free energy per unit length $F$ and
the pressure $P$ are given by
\begin{equation}
F=\mu n_c- P, \qquad P= \frac{T}{2\pi}\int_{-\infty}^{\infty}
\ln(1+e^{-\varepsilon(k)/T}) dk, \label{Pressure}
\end{equation}
where $n_c$ denotes the particle density.

\section{Low-temperature thermodynamics}
\label{therm} The most complicated  part of  the TBA equations is
the string part (which characterizes the spin excitations)
consisting of an infinite number of string functions
$\phi_{n}(\lambda)$ ((\ref{TBA-lambda}) or (\ref{TBA-phi-o}) and
(\ref{TBA-phi-n})). These coupled nonlinear integral equations have
not been solved in the most generic manner, but the obstacles can be
overcome if we make certain assumptions for the parameters involved.
Among them, one of the most crucial cases which we consider below,
is the low-temperature limit $T\ll 1$. In this limit, the TBA
equations reduce to a set of linearly coupled equations, which are
easier to deal with. Moreover for the strong coupling regime $c\gg
1$ in a weak magnetic field $H\ll 1$, we can solve the linear
integral equations {\it analytically}. Below we derive the
analytical solutions in this physical regime: strong coupling $c\gg
1$, weak magnetic field $H\ll 1$ and low-temperature $T\ll 1$. The
low-temperature thermodynamics for the generic case of $c>0$ and
$H\le H_c$ ($H_c$ is the critical field) is derived in
Appendix~\ref{low-temp}.

We first observe from equation (\ref{TBA-phi-n}) that
$\phi_{n}(\lambda)>0$ for $n>1$, because  $s(\lambda)>0$ and
$\ln(1+e^{\phi_{n}(\lambda)/T})>0$ for $\lambda\in \mathbb{R}$ and
$n\ge 1$.
This positivity condition implies that the function
$T\ln(1+e^{-\phi_{n}(\lambda)/T})\rightarrow 0$ for $T\rightarrow
0$ and $n>1$.  Therefore, in the low temperature limit $T\ll 1$,
all the higher spin string  functions drop off leaving only the function
$\phi_{1}(\lambda)$ in the first set of TBA equations. The revised
form is
\begin{align}
\varepsilon(k)&=k^{2}-\mu-\frac{H}{2}-Ta_{1}\ast\ln(1+e^{-\phi_{1}(k)/T})
\label{TBA-epsilon}\\
\phi_{1}(\lambda)&=
H-Ta_{1}\ast\ln(1+e^{-\varepsilon(\lambda)/T})+Ta_{2}\ast
\ln(1+e^{-\phi_{1}(\lambda)/T}).
\label{TBA-phi}
\end{align}

We now have to solve two coupled integral equations with only two
unknown functions $\varepsilon(k)$ and $\phi_{1}(\lambda)$. Let us
analyze them in the strong coupling regime $c\gg 1$, where the above
equations are further simplified. Analyzing the dispersion
$\varepsilon(k)$, we are able to rewrite the term \cite{GBT}
\begin{equation}
T a_1\ast\ln\left(1+e^{-\varepsilon(\lambda)/T}\right) \approx 2\pi
Pa_{1}(\lambda)+O\left(\frac{1}{c^{3}}\right), \label{appro}
\end{equation}
where the major contribution to the integral comes from a finite range
$(-k_{0},k_{0})$. As shown in Appendix~\ref{ground}, one notices
that the points $\pm k_0$ correspond to the Fermi points in the
charge Fermi sea. Thus equation (\ref{TBA-phi}) simplifies to
\begin{equation}
\phi_{1}(\lambda)=H-2\pi
Pa_{1}(\lambda)+Ta_{2}\ast\ln\left(1+e^{-\phi_{1}(\lambda)/T}\right).
\label{TBA-phi2}
\end{equation}

Using the Fourier transform, this equation and
Eq.~\eqref{TBA-epsilon} can also be expressed as
\begin{align}
&\varepsilon(k)=k^2-\mu-2\pi P K(k)-T s\ast \ln(1+e^{\phi_1(k)/T})
\nonumber  \\
&\phi_1(\lambda)=\frac{H}{2}-2\pi P s(\lambda)+T
K\ast\ln(1+e^{\phi_1(\lambda)/T}). \label{energy-spin}
\end{align}
To proceed further, let us separate the function $\phi_1(\lambda)$
into two parts:
\begin{equation}
\phi_1(\lambda)=\phi_1^{(0)}(\lambda)+\phi_1^{(1)}(\lambda).
\label{decomp}
\end{equation}
The first part $\phi_1^{(0)}(\lambda)$ corresponds to the leading
order term when $T=0$, while the second part $\phi_1^{(1)}(\lambda)$
is the first order correction to the limit $T\to 0$. Analyzing the
leading term $H/2-2\pi P s(\lambda)$ in Eq.~\eqref{energy-spin}, we
find that $\phi_1^{(0)}(\lambda)$  should satisfy the
linear integral equation
\begin{equation}
\phi^{(0)}_1(\lambda)=\frac{H}{2}-2\pi P_0
s(\lambda)+K\ast\phi_1^{(0)+}(\lambda) \label{phi-0}
\end{equation}
where $P_0$ denotes the pressure at $T=0$. Here we have divided
$\phi_1^{(0)}(\lambda)$ into its positive and negative parts:
\begin{align}
&\phi_1^{(0)}(\lambda)=\phi_1^{(0)+}(\lambda)+\phi_1^{(0)-}(\lambda), \nonumber \\
&\phi_1^{(0)-}(\lambda)=\begin{cases}
                                        \phi_1^{(0)}(\lambda) & \text{ for $|\lambda|\le \lambda_0$} \\
                                        0                                & \text{ for $|\lambda|>\lambda_0$}
                                        \end{cases}.
\label{fermi}
\end{align}

Note that the function $\phi_1^{(0)}(\lambda)$ is nothing but the
dressed energy  $\varepsilon_s(\lambda)$ \eqref{dress} denoting the
energy of a spinon excitation with rapidity $\lambda$, and the
points $\pm \lambda_0$ are the Fermi points (see
Appendix~\ref{ground}  for details). On the other hand, the function
$\varepsilon(k)$ at $T=0$ corresponds to the dressed energy
$\varepsilon_c(k)$ (\ref{dress}) describing the energy of charge
excitation with momentum $k$.

Substituting Eq.~\eqref{decomp}
into Eq.~\eqref{energy-spin}, and subtracting Eq.~\eqref{phi-0}
from the resulting equation gives
\begin{align}
\phi_{1}^{(1)}(\lambda) &=-2\pi (P-P_0)s(\lambda)+ \int_{|\mu| \ge
\lambda_0}K(\lambda-\mu)
                      \left(
                 T \ln \left( 1+e^{(\phi_{1}^{(0)}(\mu)+\phi_{1}^{(1)}(\mu))/T}\right)
                  -\phi_{1}^{(0)+}(\mu)
                       \right)
                         d\mu                \nonumber \\
&\quad +\int_{|\mu|\le \lambda_{0}}T K(\lambda-\mu)
 \ln\left(1+e^{(\phi_{1}^{(0)}(\mu)+\phi_{1}^{(1)}(\mu))/T}\right)d\mu .
\end{align}
An iteration procedure shows that $\phi_1^{(1)}(\lambda)=o(T)$.
Thus one sees that
\begin{equation}
\phi_1^{(1)}(\lambda)\approx -2\pi(P-P_0)s(\lambda)+E_K(\lambda)
           +
\int_{|\mu| \ge \lambda_0}K(\lambda-\mu) \phi_{1}^{(1)}(\mu)
                         d\mu
\end{equation}
where we denote
\begin{equation}
E_K(\lambda)=T K \ast \ln \left( 1+e^{-|\phi_{1}^{(0)}(\lambda)|/T}\right).
\end{equation}

In the limit $T\to 0$, the major contributions towards
$E_K(\lambda)$ come from the regions near $\pm\lambda_{0}$. Hence we
expand
 $\phi_{1}^{(0)}(\lambda)$ around
$\lambda = \pm\lambda_{0}$
\begin{equation}
\phi_{1}^{(0)}(\lambda)=t(\lambda-\lambda_{0})
+O((\lambda-\lambda_{0})^{2})
\end{equation}
where
$t\equiv\left.d\phi_{1}^{(0)}(\lambda)/d\lambda\right|_{\lambda=\lambda_{0}}$. Then we find
\begin{align}
\nonumber E_{K}(\lambda) &\approx T\int_{|\mu-\lambda_0|<\epsilon}
K(\lambda-\mu)\ln(1+e^{-t|\mu-\lambda_{0}|/T})d\mu
\\ &\approx \nonumber \frac{2T^{2}}{t}[K(\lambda-\lambda_{0})+
K(\lambda+\lambda_{0})]\int_{0}^{\infty}\ln\left(1+e^{-u}\right)du
\\ &= \frac{\pi^{2}T^{2}}{6t}[K(\lambda-\lambda_{0})+K(\lambda+\lambda_{0})].
\end{align}
Therefore we obtain a linear integral equation which
determines $\phi_1^{(1)}(\lambda)$, namely
\begin{equation}
\phi_{1}^{(1)}(\lambda)\approx-2\pi(P-P_0)s(\lambda)+
\frac{\pi^{2}T^{2}}{6t}[K(\lambda-\lambda_{0})+K(\lambda+\lambda_{0})]
+\int_{|\mu|\ge \lambda_0} K(\lambda-\mu)\phi_{1}^{(1)}(\mu)d\mu.
\label{phi-1}
\end{equation}

In completely the same way, one finds that the low temperature
behavior of $\varepsilon(k)$ \eqref{energy-spin} is described by
\begin{equation}
\varepsilon(k)\approx k^{2}-\mu-2\pi PK(k)-
\frac{\pi^{2}T^{2}}{6t}[s(k-\lambda_{0})+s(k+\lambda_{0})]
-\int_{|\lambda|\ge
\lambda_0}s(k-\lambda)\left[\phi_{1}^{(0)}(\lambda)+\phi_{1}^{(1)}(\lambda)\right]d\lambda.
\label{epsilon-1}
\end{equation}

Eq.~\eqref{phi-0} and Eq.~\eqref{phi-1}  can be solved for $H\ll 1$
via the Wiener-Hopf technique as in
\cite{Nepomechie1992,Nepomechie1993, book3}. For convenience, let us
introduce the functions
\begin{equation}
y^{(k)}(\lambda):=\phi_1^{(k)}(\lambda+\lambda_0) \qquad (k=0,1).
\label{y-func}
\end{equation}
By definition (see Eq.~\eqref{fermi}), the Fermi points $\pm
\lambda_0$ are determined by the condition
\begin{equation}
\phi_1^{(0)}(\lambda_0)=y^{(0)}(0)=0.
\label{condition-fermi}
\end{equation}

Applying the iterative procedure to Eq.~\eqref{phi-0}, one sees
$y^{(0)}(0)=0=H/2-2\pi P_0 s(\lambda_0)+O(H)$. Solving this
equation, one finds that $\lambda_0\approx -\ln H$ for $H\ll 1$.
Because $K(\lambda)$ rapidly decreases with $\lambda>0$, we may
solve the integral equations \eqref{phi-0} and \eqref{phi-1} by
expanding
\begin{equation}
y^{(k)}(\lambda)=\sum_{n=0}^{\infty} y^{(k)}_n(\lambda) \qquad
(k=0,1).
\end{equation}
$y_n^{(k)}(\lambda)$ obeys the integral equation
\begin{equation}
y_n^{(k)}(\lambda)=g^{(k)}_n(\lambda)+\int_0^\infty
K(\lambda-\mu)y_n^{(k)}(\mu) d\mu \label{scaling}
\end{equation}
where the driving terms $g_n^{(0)}(\lambda)$ and
$g_n^{(1)}(\lambda)$ in the limit of $T\ll 1$ are explicitly given
by
\begin{align}
&g_0^{(0)}(\lambda)=\frac{H}{2}-2\pi P_0 s(\lambda+\lambda_0) \quad
\nonumber \\
&g_0^{(1)}(\lambda)=-2\pi(P-P_{0})s(\lambda+\lambda_{0})+\frac{\pi^2
T^2}{6t}K(\lambda) \nonumber \\
&g_n^{(k)}(\lambda)=\int_0^\infty
K(\lambda+\mu+2\lambda_0)y^{(k)}_{n-1}(\mu) d\mu \qquad (n \ge 1).
\end{align}

The above equations are the so-called Wiener-Hopf type integral
equations. In Appendix~\ref{WH}, a method to solve the Wiener-Hopf
type integral equations is given. Let us decompose
$\widehat{y}^{(k)}(\omega)$ into a sum of two parts, i.e.,
$\widehat{y}^{(k)}(\omega)=\widehat{y}^{(k)}_+(\omega)+\widehat{y}^{(k)}_-(\omega)$,
where $\widehat{y}_+^{(k)}(\omega)$ ($\widehat{y}_-^{(k)}(\omega)$)
is analytic in the upper (lower) half plane (see
Eq.~\eqref{decomp-y} in Appendix~\ref{WH}). For the leading terms
$\widehat{y}_0^{(k)}(\omega)$, we obtain  (see Eqs~\eqref{y0} and \eqref{y1})
\begin{align}
&\widehat{y}_{0+}^{(0)}(\omega)=G_+(\omega)
\left[
\frac{i H G_-(-i\epsilon)}{2(\omega+i\epsilon)}-\frac{2\pi i P_0}{c}
\frac{G_-(-\pi i/c)e^{-\pi \lambda_0/c}}{\omega+\pi i/c}
\right]+O(H^2)  \nonumber \\
&\widehat{y}_{0+}^{(1)}(\omega)=  \frac{\pi^2
T^2}{6t}(G_+(\omega)-1)-\frac{2\pi
i(P-P_{0})}{c}\frac{G_{+}(\omega)G_{-}(-\pi
i/c)e^{-\pi\lambda_{0}/c}}{\omega+\pi i/c} \label{y-hat}
\end{align}
where $G_\pm(\omega)$ is defined by Eq.~\eqref{g-func0}. Using the
formula as in \eqref{formula-y0}, and combining the above equation
with the condition \eqref{condition-fermi}, we determine the
leading term of the Fermi points $\lambda_0$ to be
\begin{equation}
\lambda_0\approx\frac{c}{\pi}\ln\left(\frac{H_0}{H}\right), \qquad
H_0=\frac{4\pi P_0G_-(-\pi i/c)}{cG_-(0)}=\frac{4\pi
P_0}{c}\sqrt{\frac{\pi}{2 e}} =\sqrt{\frac{\pi^3}{2
e}}H_c+O\left(\frac{1}{c^{2}}\right) \label{fermi-point}
\end{equation}
where we have used the formulae \eqref{factorization} and
\eqref{g-func}. To derive the last equality in the second equation we used
the property (see Appendix~\ref{ground})
\begin{equation}
\frac{P_0}{c}=-\frac{1}{2\pi c}\int_{-k_0}^{k_0}\varepsilon_c(k) dk
\approx \frac{H_c}{4}+O\left(\frac{1}{c}\right)
\end{equation}
where $H_c$ is the critical magnetic field, where all fermion spins
point up (see \cite{Guan2008} or Eq.~\eqref{Hc} in
Appendix~\ref{ground}), with $H_c\approx 8 n_c^3\pi^2/3c$.

From the relation \eqref{y-func}, the integral in \eqref{epsilon-1}
can be evaluated as follows:
\begin{align}
\int_{|\lambda|\ge
\lambda_0}s(k-\lambda)\phi_{1}^{(k)}(\lambda)d\lambda &\approx
\frac{2e^{-\frac{\pi}{c}\lambda_0}}{c}\int_{0}^{\infty}e^{-\frac{\pi}{c}\lambda}
y_0^{(k)}(\lambda)d\lambda  \nonumber \\
&=\frac{2e^{-\frac{\pi}{c}\lambda_0}}{c}\int_{0}^{\infty}e^{-\frac{\pi}{c}\lambda}
d\lambda
\int_{-\infty}^{\infty}\frac{d\omega}{2\pi}e^{-i\lambda\omega}
\widehat{ y}_{0+}^{(k)}(\omega)  \nonumber \\
&=-\frac{ie^{-\frac{\pi}{c}\lambda_0}}{\pi c}
\int_{-\infty}^{\infty}\frac{d\omega}{\omega-\pi i/c}
\widehat{ y}_{0+}^{(k)}(\omega)  \nonumber \\
&=\frac{2e^{-\frac{\pi}{c}\lambda_0}}{c}\widehat{y}_{0+}^{(k)}
\left(\frac{\pi i}{c}\right) \label{y-int}
\end{align}
where the relation $s(k\pm|\lambda+\lambda_0|)\approx
e^{-\pi(\lambda+\lambda_0)/c}/c$ is used in the first line.
Substitution of Eq.~\eqref{y-int} together with
Eqs.~\eqref{fermi-point}, \eqref{y-hat} and \eqref{g-func} into
Eq.~\eqref{epsilon-1} yields
\begin{equation}
\varepsilon(k)\approx k^2-\mu-2\pi P K(k)-\frac{cPH^2}{4\pi^2
P_0^{2}}- \frac{\pi T^2 H}{6\sqrt{2}P_0 t}.
\end{equation}
Inserting the relations
\begin{equation}
K(k)\approx \frac{\ln 2}{c\pi}, \qquad t\approx
y_0'(0)=-\lim_{\omega\to\infty}\omega^2
\widehat{y}_{0+}^{(0)}(\omega) =\frac{\pi H}{\sqrt{2}c} \label{K}
\end{equation}
and using $P\approx P_0$, we arrive at
\begin{equation}
\varepsilon(k)\approx k^2-\mu-\frac{2 P\ln{2}}{c}-\frac{c H^2}{4\pi
^2 P}- \frac{c T^2 }{6P}=k^2-A \label{denergy}
\end{equation}
where $A:=\mu+2P\ln 2/c+cH^{2}/4\pi^{2}P+cT^{2}/6P$.
We would like to address that the calculation of the dressed energy potential (\ref{denergy}) is essential for catching spin-charge separation signature and quantum criticality at low temperatures. The terms in the function $A$ show  an  important implementation of spin density and charge density fluctuations that reveals a physical origin of spin-charge separation.  This is clearly seen from the following low temperature and finite temperature thermodynamics.  In view of this validity of catching  this  universal nature,  we see that the result (\ref{denergy}) is also helpful to   numerics.  Using this function $A$, we then
perform integration by parts on the pressure of the system
(\ref{Pressure}) to get
\begin{equation}
P=
\frac{1}{\pi}\int_{0}^{\infty}\frac{\sqrt{\varepsilon}d\varepsilon}
                                                    {1+e^{(\varepsilon-A)/T}}=
-\frac{1}{\sqrt{4\pi}}T^{\frac{3}{2}}
\mathrm{Li}_{\frac{3}{2}}\left(-e^{A/T}\right) \label{pressure-pl}
\end{equation}
where $\mathrm{Li}_s(z)$ is polylogarithm function defined by
\begin{equation}
\mathrm{Li}_s(z)=\sum_{k=1}^{\infty}\frac{z^{k}}{k^{s}}.
\end{equation}

For fixed particle density, the chemical potential is determined by
solving the equation $n_c=\partial P/\partial \mu$.  By inserting
Eq.~\eqref{pressure-pl} into Eq.~\eqref{Pressure}, the
low-temperature thermodynamics in the region $H\ll 1$ and $c\gg 1$
is completely determined. To capture universal features of the
low-temperature thermodynamics, we further expand
Eq.~\eqref{pressure-pl} by making use of Sommerfeld's lemma (see
\cite{Pathria,GBT} for instance):
\begin{equation}
P=\frac{2A^{3/2}}{3\pi}\left(1+\frac{\pi^{2}}{8}\left(\frac{T}{A}\right)^{2}
+O(T^{4})\right).\label{Pressure-Sommerfeld}
\end{equation}

\begin{figure}[t]
{{\includegraphics [width=0.90\linewidth]{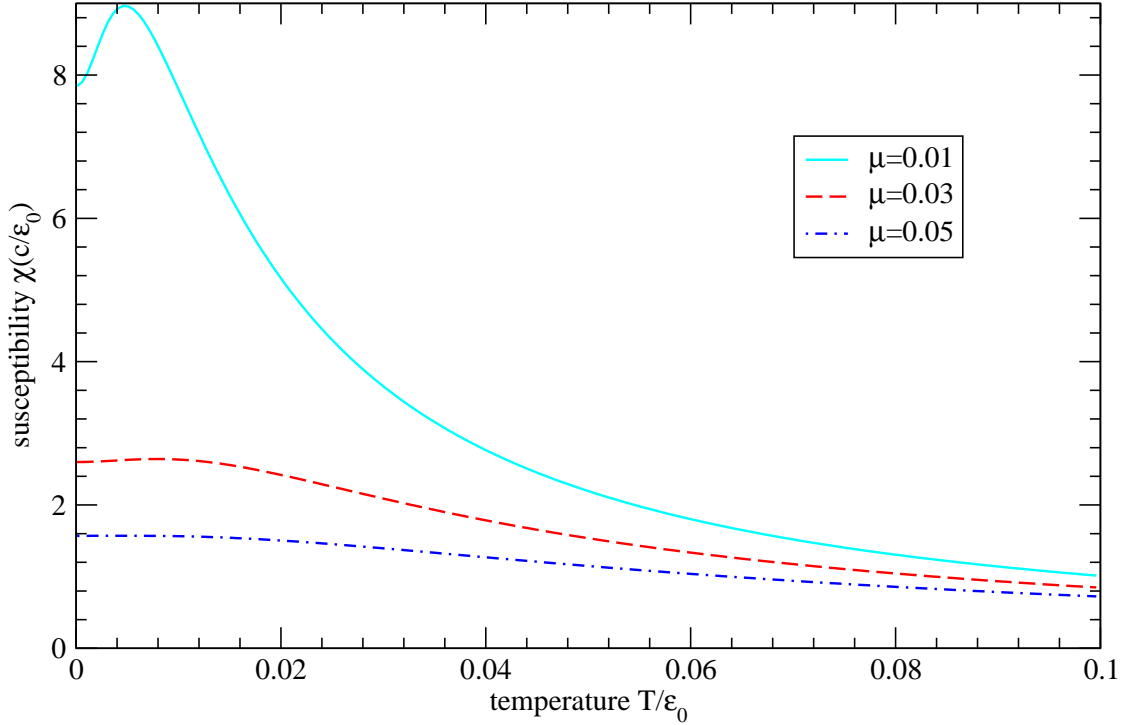}}}
\caption{Susceptibility vs temperature for different values of chemical potentials   $\mu=0.01,0.03,0.05$
   and $h=0.001$. At $T=0$, the susceptibility
   values for different values of chemical potentials are consistent with the field theory  prediction $\chi
   v_s=\theta/\pi$. This possibly gives a way to test
   effective spin velocities   of the  spin-1/2 ultracold atoms with a repulsive delta-function interaction. 
 }\label{fig:susceptibility-R}
\end{figure}

Furthermore, we use the relation $n_c=\partial P/\partial\mu$ and
repeatedly iterate the terms to eliminate those with orders higher
than $T^{2}$, $H^{2}$ and $1/c$. After some lengthy algebra, we finally
obtain the chemical potential
\begin{equation}
\mu \approx \pi^{2}n_c^{2} \left(1-\frac{16\ln 2}{3\gamma}\right)+
\frac{3\gamma H^{2}}{4\pi^{4}n_c^{2}} \left(1+\frac{3\ln
2}{\gamma}\right)+ \frac{\gamma T^{2}}{2\pi^{2}n_c^{2}}\left(1+
\frac{3\ln
2}{\gamma}\right)+\frac{T^{2}}{12n_c^{2}}\label{mu-Sommerfeld}
\end{equation}
where $\gamma$ denotes $\gamma:=c/n_c$. Substituting
(\ref{mu-Sommerfeld}) into the expression for the pressure
(\ref{Pressure-Sommerfeld}) and then iterating the pressure itself,
we obtain
\begin{equation}
P\approx \frac{2}{3}\pi^{2}n_c^{3}\left(1-\frac{6\ln
2}{\gamma}\right)+\frac{9\gamma
H^{2}}{8\pi^{4}n_c}\left(1+\frac{4\ln
2}{\gamma}\right)+\frac{3\gamma
T^{2}}{4\pi^{2}n_c}\left(1+\frac{4\ln
2}{\gamma}\right)+\frac{T^{2}}{6n_c}. \label{Pressure-s}
\end{equation}
The free energy of the system defined by Eq.~\eqref{pressure} is
given by
\begin{equation}
F\approx\frac{1}{3}\pi^{2}n_c^{3}\left(1-\frac{4\ln
2}{\gamma}\right)-\frac{3\gamma
H^{2}}{8\pi^{4}n_c}\left(1+\frac{6\ln 2}{\gamma}\right)-\frac{\gamma
T^{2}}{4\pi^{2}n_c}\left(1+\frac{6\ln
2}{\gamma}\right)-\frac{T^{2}}{12n_c}.\label{Free-E}
\end{equation}
From this free energy \eqref{Free-E}, we have the susceptibility
\begin{equation}
\chi =-\frac{\partial^2 F}{\partial H^2}\approx
\frac{3\gamma}{4\pi^{4}n_c} \left(1+\frac{6\ln 2}{\gamma}\right).
\end{equation}
This susceptibility can be  possibly tested in an trapped 1D Fermi gas of cold atom.  We rewrite the pressure (\ref{pressure-pl}) 
\begin{equation}
p=-\sqrt{\frac{m}{2\pi\hbar^2}}T^{\frac{3}{2}}\mathrm{Li}_{\frac{3}{2}}\left(-e^{X}\right)\label{Pres-Rep-L}
\end{equation}
where
\begin{eqnarray}
X&=&\frac{\mu}{T}-\frac{\ln 2}{\sqrt{\pi}}\sqrt{\frac{T}{\varepsilon_0}}\mathrm{Li}_{\frac{3}{2}}\left(-e^{\frac{\mu}{T}}\right)-\frac{1}{2\pi^{\frac{3}{2}}}\frac{H^2}{T^2}\frac{1}{\sqrt{\frac{T}{\varepsilon_0}}\mathrm{Li}_{\frac{3}{2}}\left(-e^{\frac{\mu}{T}}\right)}\nonumber\\
&& -\frac{\sqrt{\pi}}{3}\frac{1}{\sqrt{\frac{T}{\varepsilon_0}}\mathrm{Li}_{\frac{3}{2}}\left(-e^{\frac{\mu}{T}}\right)}.
\end{eqnarray}
In dimensionless units, i.e. $h=H/\varepsilon_0$ and $\mu=\mu/\varepsilon_0$ with $\varepsilon_0=\frac{\hbar^2}{2m}c^2$, the susceptibility is given by
\begin{equation}
\frac{\chi}{\varepsilon_0} =\frac{c}{\varepsilon_0}\frac{3\gamma}{4\pi^4 n'(\mu,h)}\left(1+\frac{6\ln 2}{\gamma}\right).
\end{equation}
which agrees with the  field theory prediction  $\chi v_s=\theta/\pi$. Here $\theta=1/2$.  The holon velocity $v_c$ and spinon velocity
$v_s$ at $H=0$  are given in  \eqref{fermi-h0} in Appendix~\ref{low-temp}. In the above equation, the density is given by 
\begin{eqnarray}
n':&=& \frac{n}{\sqrt{\varepsilon_0}} \approx \frac{\mu}{\pi}\left[1-\frac{3}{8\pi\sqrt{\mu}}\left(\frac{h}{\mu}\right)^2\left(1+\frac{3\sqrt{\mu}\ln2}{\pi}\right)+\frac{8\sqrt{\mu}\ln2}{3\pi}\right]
\end{eqnarray}
and the dimensionless interaction strength is given by 
\begin{eqnarray}
\frac{1}{\gamma}\approx \frac{\sqrt{\mu}}{\pi}-\frac{3}{8\pi^2}\left(\frac{h}{\mu}\right)^2.
\end{eqnarray}
In Fig. \ref{fig:susceptibility-R}, we plot the susceptibility for different values of the chemical potentials at low temperatures. 
In contrast,  in Chapter 13 of Ref. \cite{book3}, the  thermal and magnetic properties of the 1D Hubbard model   
are  plotted by using the result obtained for  the thermodynamics of the 1D Hubbard model by  numerically solving 
nonlinear integral equations (NLIE). The thermodynamics for the Hubbard model can be calculated by the quantum transfer matrix method 
for all temperatures. 
Here we have obtained an explicit low temperature expansion for the fermion model.  The result  (\ref{Pres-Rep-L}) for the 
pressure  contains  the  spin density and charge thermal potentials  at  criticality and may thus  possibly be used to test the 
spin and charge velocities in experiments with ultracold atomic fermions  in a 1D  harmonic trap. 

\begin{figure}[ht]
\includegraphics[width=.8\columnwidth]{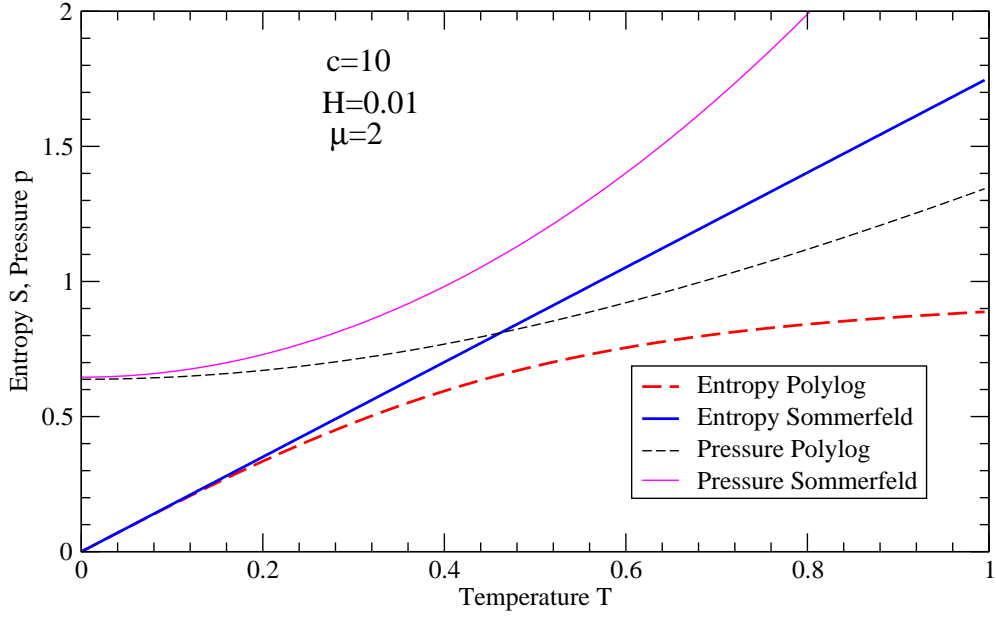}
\caption{(Color online) Pressure and entropy vs temperature $T$. Solid
lines show the pressure \eqref{Pressure-s} and entropy obtained from
the free energy \eqref{free-E-s}; dashed lines show pressure and
entropy obtained from the polylogarithm function \eqref{pressure-pl}
which gives the precise thermodynamics for temperatures below the
Fermi temperature
$k_{B}T_{F}=E_{F}=\frac{\pi^{2}\hbar^{2}n^{2}}{6m}$ in the strong
coupling limit. At low temperatures, they are indeed consistent with
the field theory predictions. The deviation of the entropy from the
linear temperature dependence marks a breakdown of the two massless
field theories.}\label{fig:S-C-Sep}
\end{figure}

By substitution of the holon velocity $v_c$ and spinon velocity
$v_s$ at $H=0$ (see \eqref{fermi-h0} in Appendix~\ref{low-temp}),
the free energy (\ref{Free-E}) suggests the universal low temperature
form of spin-charge separation theory, namely
\begin{equation}
F= E_0-\frac{\pi C T^2}{6}\left( \frac{1}{v_s} +\frac{1}{v_c}
\right), \label{free-E-s}
\end{equation}
where $E_0$ denotes the ground state energy density and $C=1$. The
above behavior can also be derived for generic $c$ and $H$ (see
Appendix~\ref{low-temp}). This expression corresponds to two central
charge $C=1$ conformal field theories \cite{Affleck1986}. This
universal nature of  (\ref{free-E-s}) means that the low-lying excitations
are decoupled into two massless degrees of freedom which are
described by two Gaussian theories. However, if the excitations
involve highly excited states, these theories break down. In Fig.~
\ref{fig:S-C-Sep} we compare the exact thermodynamics with  the
predictions of conformal field theory and show the breakdown of the
two Gaussian theories at higher temperatures.

\section{Dressed charge}\label{section-dressed}
The excitation spectrum for repulsive spin-1/2 fermions is gapless
for any magnetic field strength $H$. As a consequence, the
asymptotic behavior of the correlation functions of the model can be
described by conformal field theory (CFT)
\cite{Cardy1986,Blote1986,Affleck1986}. CFT relates the critical
exponents of the correlation functions to the finite-size
corrections in the energy spectrum. The basic tool used
to determine the critical exponents from the finite-size corrections
is the dressed charge matrix. For this model, it is a $2\times 2$
matrix which couples the charge and spin degrees of freedom together
and at the same time governs the excitations of charge and spin
waves near the Fermi surface.

The dressed charge matrix of this system is explicitly given by
\begin{equation}
\mathbf{Z}=\left(
             \begin{array}{cc}
               Z_{cc}(k_{0}) & Z_{cs}(\lambda_{0}) \\
               Z_{sc}(k_{0}) & Z_{ss}(\lambda_{0}) \\
             \end{array}
           \right)
\end{equation}
while the integral equations of its elements are given by
\begin{eqnarray}
Z_{cc}(k) &=&
1+\int_{-\lambda_{0}}^{\lambda_{0}}a_{1}(k-\lambda)Z_{cs}(\lambda)d\lambda
\label{D1}
\\ Z_{cs}(\lambda) &=&
\int_{-k_{0}}^{k_{0}}a_{1}(\lambda-k)Z_{cc}(k)dk-\int_{-\lambda_{0}}^{\lambda_{0}}a_{2}(\lambda-\mu)Z_{cs}(\mu)d\mu
\label{D2}
\\ Z_{sc}(k) &=&
\int_{-\lambda_{0}}^{\lambda_{0}}a_{1}(k-\lambda)Z_{ss}(\lambda)d\lambda
\label{D3}
\\ Z_{ss}(\lambda) &=&
1+\int_{-k_{0}}^{k_{0}}a_{1}(\lambda-k)Z_{sc}(k)dk-\int_{-\lambda_{0}}^{\lambda_{0}}a_{2}(\lambda-\mu)Z_{ss}(\mu)d\mu.
\label{D4}
\end{eqnarray}

\begin{figure}
\centering
\begin{tabular}{cc}
\epsfig{file=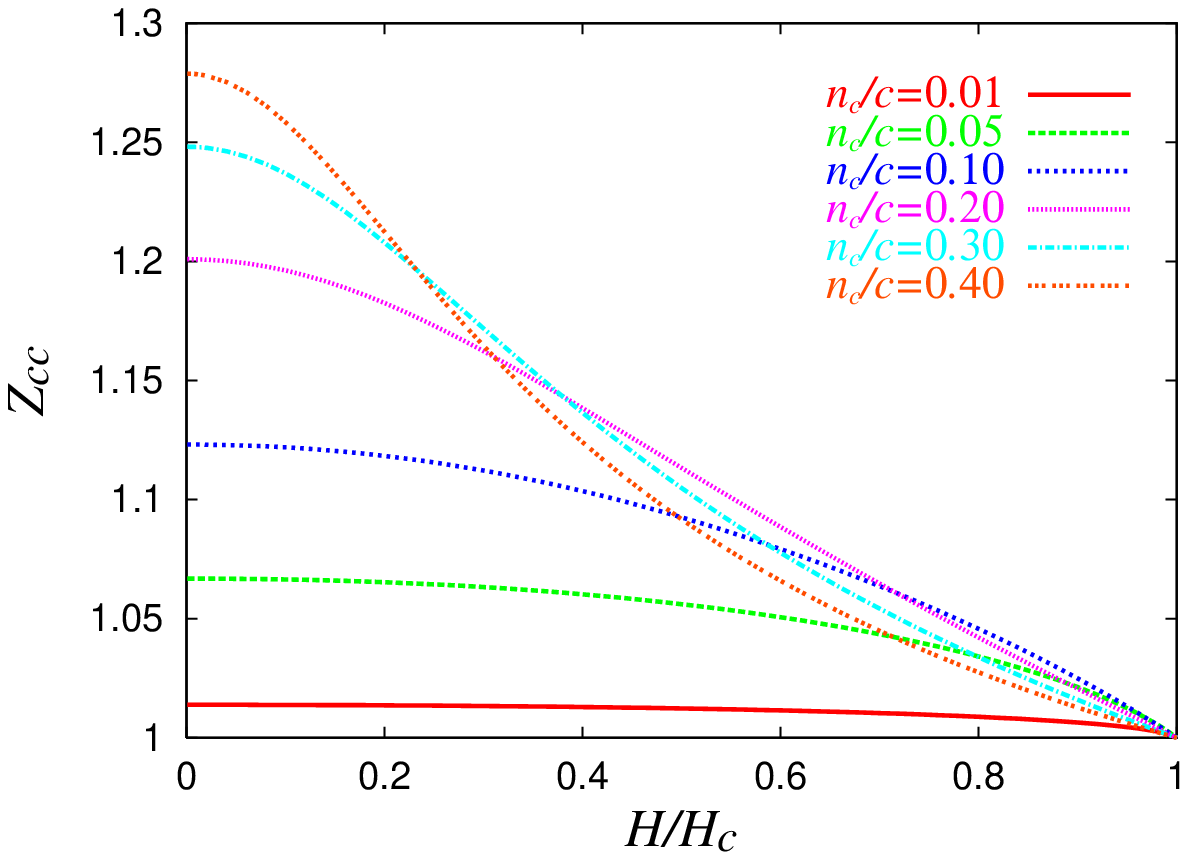,width=0.5\linewidth,clip=} &
\epsfig{file=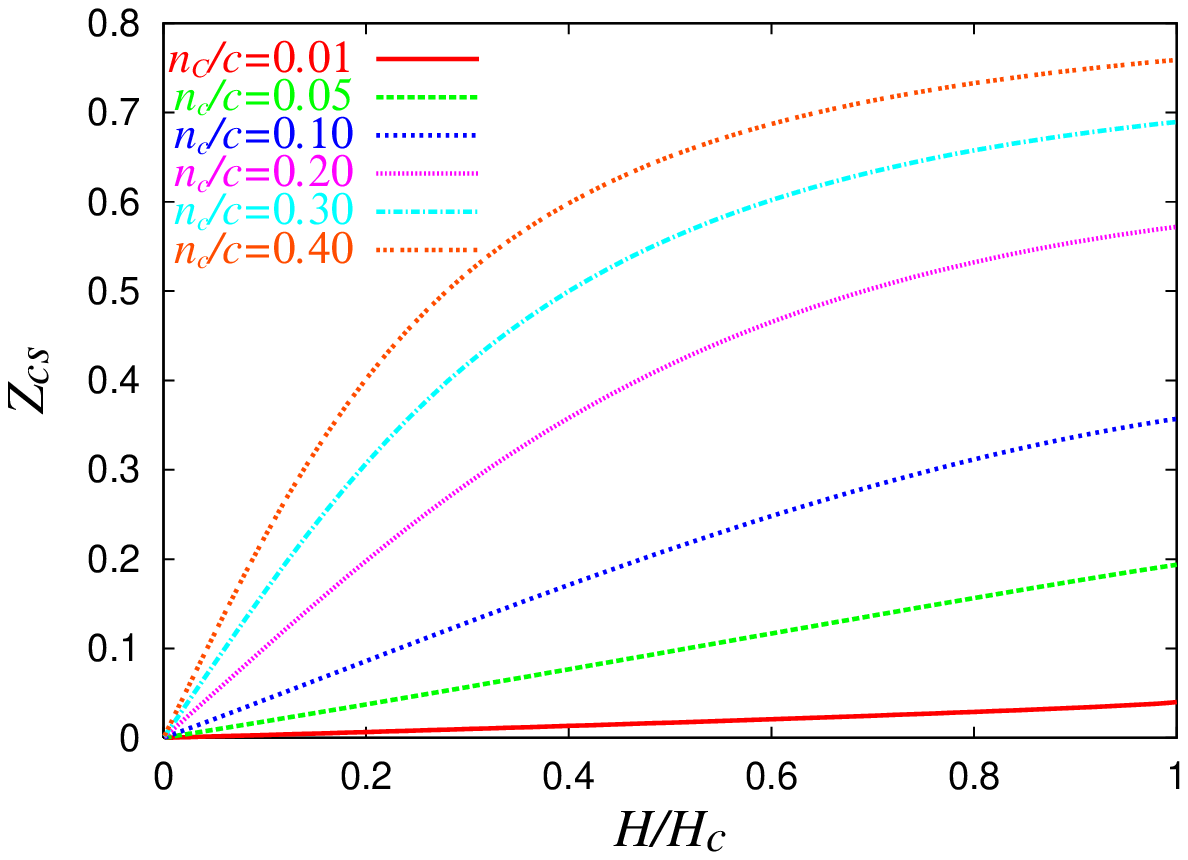,width=0.5\linewidth,clip=} \\
\epsfig{file=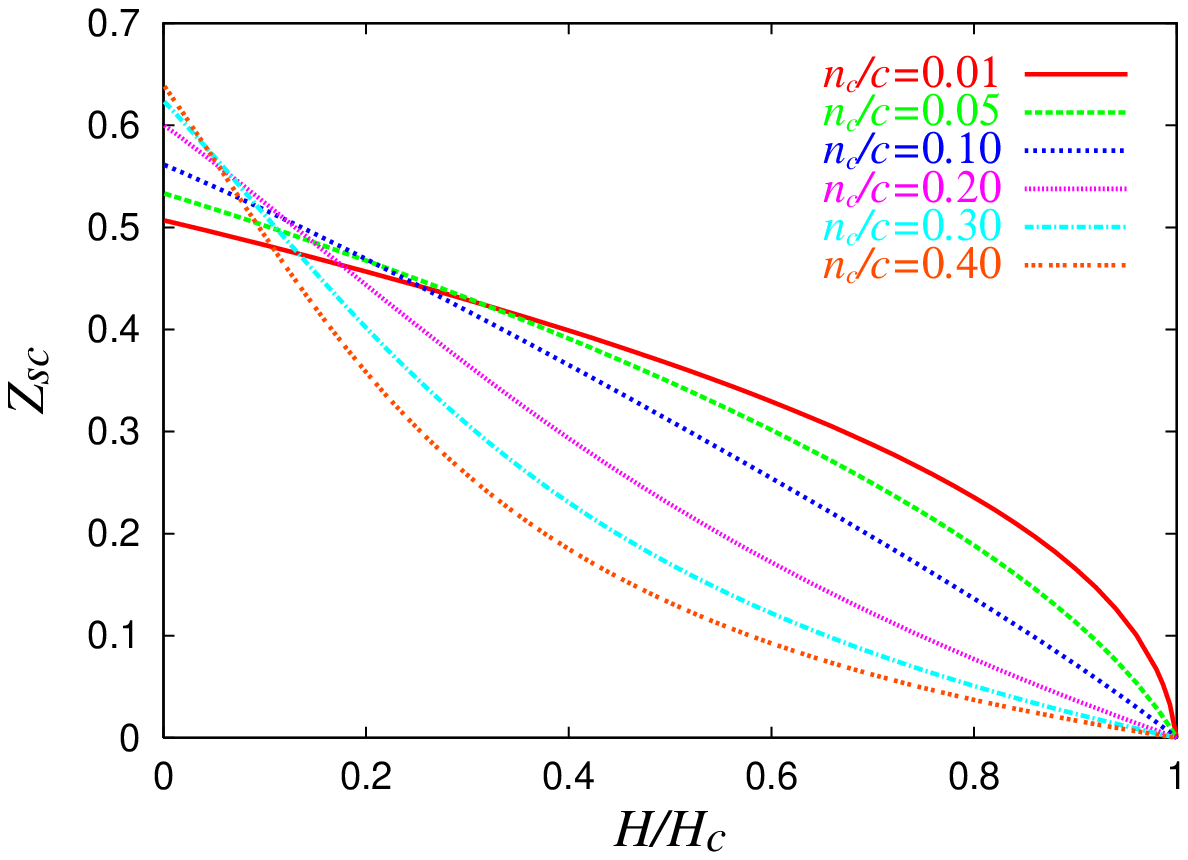,width=0.5\linewidth,clip=} &
\epsfig{file=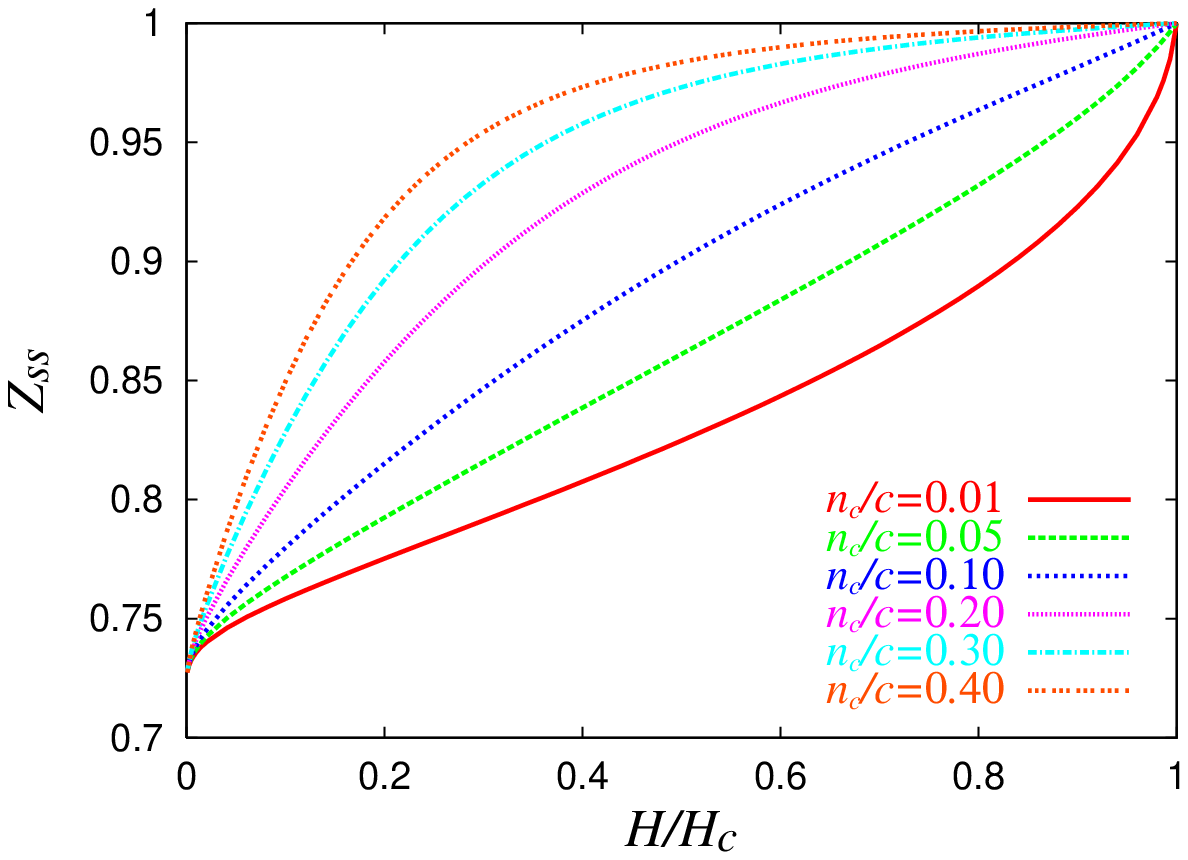,width=0.5\linewidth,clip=}
\end{tabular}\caption{(Color online) This figure shows
the dressed charges $Z_{cc}(k_{0})$, $Z_{cs}(\lambda_{0})$,
$Z_{sc}(k_{0})$ and $Z_{ss}(\lambda_{0})$ as a function of the
external field $H/H_{c}$. The dressed charges are plotted for
different values of $n_{c}/c$ which is the inverse of the
interaction parameter $\gamma$. These curves are plotted by
numerically solving Eqs.~\eqref{D1}--\eqref{D4}.}\label{fig:Z}
\end{figure}

This set of equations are in turn made up of two coupled sets of
equations. Equations (\ref{D1}) and (\ref{D2}) can be treated
separately from equations (\ref{D3}) and (\ref{D4}). The following
relations are useful for further calculation:
\begin{align}
\int_{-k_{0}}^{k_{0}}\frac{Z_{sc}(k)}{2\pi}dk &=
\int_{-\lambda_{0}}^{\lambda_{0}}\frac{Z_{cs}(\lambda)}{2\pi}d\lambda=
\int_{-\lambda_{0}}^{\lambda_{0}}\rho_{s}(\lambda)d\lambda=n_{\downarrow} \label{equiv1} \\
\int_{-k_{0}}^{k_{0}}\frac{Z_{cc}(k)}{2\pi}dk &=
\int_{-k_{0}}^{k_{0}}\rho_{c}(k)dk=n_c \label{equiv2}
\end{align}
where $n_{\downarrow}$ is the density of down-spin fermions. To
derive these relations, we first multiply each set of density
equations with the dressed charge equations and integrate them.
While making use of the fact that the kernels are symmetric, we then
subtract one equation from the other to eliminate the same terms.
FIG.~\ref{fig:Z} shows a depiction of the numerical solutions to
Eqs.~\eqref{D1}--\eqref{D4}.

\subsection{The limit $H\ll 1$}
When $H=0$, the ground state of the system is antiferromagnetic. We
now solve the dressed charge equations under a small magnetic field
by the Wiener-Hopf technique. Considering only terms of up to order
$1/c$ in the strong coupling limit, equation (\ref{D4}) can be
written as
\begin{align}
\nonumber Z_{ss}(\lambda) &=
1+a_{1}(\lambda)\int_{-k_{0}}^{k_{0}}Z_{sc}(k)dk-\int_{-\lambda_{0}}^{\lambda_{0}}a_{2}(\lambda-\mu)Z_{ss}(\mu)d\mu
\\ &=1+2\pi
n_{\downarrow} a_{1}(\lambda)-\int_{-\lambda_{0}}^{\lambda_{0}}
a_{2}(\lambda-\mu)Z_{ss}(\mu)d\mu
\end{align}
where the property \eqref{equiv1} was used. Applying the Fourier
transform, we obtain
\begin{equation}
Z_{ss}(\lambda)=\frac{1}{2}+2\pi
n_{\downarrow} s(\lambda)+\int_{|\mu|\ge \lambda_{0}}K(\lambda-\mu)Z_{ss}(\mu)d\mu.
\label{Zss-f}
\end{equation}

This equation is the same as Eq.~\eqref{phi-0} other than the
driving term, and hence a similar procedure as introduced in
Section \ref{therm} is applicable. Introducing the function
$y(\lambda)=Z_{ss}(\lambda+\lambda_0)$ and expanding it as
$y(\lambda)=\sum_{n=0}^{\infty}y_n(\lambda)$, we obtain the
integral equation (see the corresponding equation
\eqref{scaling} in Sec.~\ref{therm})
\begin{align}
y_n(\lambda)=g_n(\lambda)+\int_0^\infty K(\lambda-\mu)y_n(\mu) d\mu
\label{scaling-2}
\end{align}
where $g_n(\lambda)$ is given by
\begin{align}
&g_0(\lambda)=\frac{1}{2}+2\pi n_{\downarrow}
s(\lambda+\lambda_{0}), \nonumber \\
&g_n(\lambda)=\int_0^\infty K(\lambda+\mu+2\lambda_0)y_{n-1}(\mu)
d\mu \qquad (n\ge 1).
\end{align}

For $n=0$, setting $a=1/2$ and $b=2\pi n_{\downarrow}$ in
Eq.~\eqref{y0},  one has
\begin{equation}
\widehat{y}_{0+}(\omega)=G_+(\omega)
\left[
\frac{i G_-(-i\epsilon)}{2(\omega+i\epsilon)}+
\frac{ 2\pi i n_{\downarrow} G_-(-\pi i/c)e^{-\pi \lambda_0/c}}{c(\omega+\pi i/c)}
\right]+O(H^2).
\label{Zss-y0}
\end{equation}
Combining this result with Eq.~\eqref{formula-y0}, and finally
substituting the relations \eqref{g-func}, \eqref{factorization} and
\eqref{fermi-point}, one obtains the first order contribution to
$Z_{ss}(\lambda_0)$:
\begin{equation}
y_0(0)=\frac{1}{\sqrt{2}}\left(1+\frac{4 n_{\downarrow}}{c}\frac{H}{H_c}\right)+O(H^2).
\label{Zss-1}
\end{equation}

To obtain the second order correction to $y(0)=Z_{ss}(\lambda_0)$,
we  must consider the contribution of $y_{1}(0)$.  The Fourier
transform of $g_1(\lambda)$ is given by
\begin{equation}
\widehat{g}_1(\omega)=e^{-2 i \lambda_0 \omega}\widehat{y}_{0+}(-\omega)
\widehat{K}(\omega)=e^{-2 i \lambda_0 \omega}\widehat{y}_{0+}(-\omega)
\left[1-\frac{1}{G_+(\omega)G_-(\omega)}\right].
\end{equation}
Here we have used the decomposition of the kernel
\eqref{factorization}. As demonstrated in Appendix.~\ref{WH}, let us
decompose $\widehat{g}_1(\omega)G_-(\omega)$ into the two parts
$\Phi_{\pm}(\omega)$ which are analytic in the upper and lower half
planes, respectively:
$\widehat{g}_1(\omega)G_-(\omega)=\Phi_+(\omega)+\Phi_-(\omega)$.
Now $\Phi_+(\omega)$ is given by
\begin{equation}
\Phi_+(\omega)= -\frac{1}{2\pi i}\int_{-\infty}^{\infty}\frac{e^{-2i
\lambda_0 x}\widehat{y}_{0+}(-x)}
{x-\omega-i\epsilon}\frac{1}{G_+(x)} d x
\end{equation}
where $\epsilon$ is a small positive constant. Note that the
function $G_+(x)$ has a branch cut along the negative imaginary
axis. Deforming the integration contour to avoid the branch cut, we
have
\begin{eqnarray}
\Phi_+(\omega)&=&\frac{1}{2\pi i}\int_0^\infty \frac{e^{-2 \lambda_0
x} \widehat{y}_{0+}(i x)}{x-i\omega} \left[\frac{1}{G_+(-i
x-\epsilon)}-\frac{1}{G_+(-i x+\epsilon)}\right]dx \nonumber \\
&=&\frac{1}{(2\pi)^{\frac{3}{2}}i}\int_0^\infty \frac{e^{-2
\lambda_0 x} \widehat{y}_{0+}(ix)}{x-i\omega}\left(\frac{cx}{2\pi
e}\right)^{\frac{cx}{2\pi}}\Gamma\left(\frac{1}{2}-\frac{cx}{2\pi}\right)(e^{\frac{icx}{2}}-e^{\frac{-icx}{2}})dx
\nonumber \\ &=&\frac{2}{(2\pi)^{\frac{3}{2}}}\int_0^\infty
\frac{e^{-2 \lambda_0 x}
\widehat{y}_{0+}(ix)}{x-i\omega}\left(\frac{cx}{2\pi
e}\right)^{\frac{cx}{2\pi}}\Gamma\left(\frac{1}{2}-\frac{cx}{2\pi}\right)
\sin\left(\frac{cx}{2}\right)dx \nonumber \\
&\approx& \frac{1}{2\pi}\int_0^\infty \frac{e^{-2 \lambda_0
x}}{-i\omega} \left(\frac{c}{\sqrt{2}}+O(x)\right)dx
\nonumber \\
&=&\frac{1}{-i\omega}
\left(\frac{c}{4\sqrt{2}\pi\lambda_0}+O\left(\frac{1}{\lambda_0^2}\right)\right).
\end{eqnarray}

Note that, from the third to the fourth step in the above equation,
we have used the fact that the integrand rapidly decreases for $x>0$
because $\lambda_0\gg 1$, and hence the integral can be
approximated by expanding the terms other than $\exp(-2\lambda_0
x)$ around $x=0$. By a relation similar to Eq.~\eqref{decomp-Phi},
$\widehat{y}_{1+}(\omega)$ is expressed as
\begin{equation}
\widehat{y}_{1+}(\omega)=G_+(\omega)\Phi_+(\omega)
=\frac{G_+(\omega)}{-i\omega}
\left(\frac{c}{4\sqrt{2}\pi\lambda_0}+O\left(\frac{1}{\lambda_0^2}\right)\right).
\end{equation}
Insertion of the relation \eqref{formula-y0} and
Eq.~\eqref{factorization} yields
\begin{equation}
y_1(0)=\frac{c}{4\sqrt{2}\pi
\lambda_0}+O\left(\frac{1}{\lambda_0^2}\right). \label{Zss-2}
\end{equation}
Therefore from \eqref{Zss-1}, \eqref{Zss-2} and \eqref{fermi-point},
we finally obtain the expression
\begin{equation}
Z_{ss}(\lambda_{0})=\frac{1}{\sqrt{2}}\left[1+
\frac{4n_{\downarrow}}{c}\left(\frac{H}{H_{c}}\right)+\frac{1}{4\ln(H_{0}/H)}\right]+O\left(\frac{1}{(\ln
H_0/H)^2}\right). \label{zss}
\end{equation}

Next we evaluate the dressed charge $Z_{sc}(k_0)$. The Fourier
transforms of \eqref{D3} and \eqref{Zss-f} give
\begin{equation}
Z_{sc}(k)=\frac{1}{2}+2\pi n_{\downarrow} K(k) -\int_{|\mu|\ge
\lambda_0}s(k-\lambda)Z_{ss}(\lambda)d\lambda.
\end{equation}
Applying the same procedure as used in the derivation of
Eq.~\eqref{y-int}, we immediately obtain
\begin{align}
Z_{sc}(k_0)&=\frac{1}{2}+2\pi n_{\downarrow} K(k_0)-
\frac{2e^{-\frac{\pi}{c}\lambda_0}}{c}\widehat{ y}_{0+}
\left(\frac{\pi i}{c}\right)  \nonumber \\
&= \frac{1}{2}+\frac{2 n_{\downarrow} \ln
2}{c}-\frac{2}{\pi^2}\left(\frac{H}{H_c}\right)+ O\left(\frac{H}{H_c
\ln (H_0/H)}\right) \label{zsc}
\end{align}
where we have used the relation \eqref{K}.

Repeating this whole process to evaluate $Z_{cs}(\lambda_{0})$ and
$Z_{cc}(k_{0})$, we find that
\begin{equation}
Z_{cs}(\lambda_{0})=\frac{2\sqrt{2}n_c}{c}
\left(\frac{H}{H_{c}}\right)+O\left(\frac{H}{H_c(\ln(H_0/H))^{2}}\right)
\end{equation}
and
\begin{equation}
Z_{cc}(k_{0})=
1+\frac{2n_c\ln 2}{c}-
\frac{4n_c}{\pi^{2}c}\left(\frac{H}{H_{c}}\right)^{2}+
O\left(\frac{H^2}{H^2_c(\ln(H_0/H))^{2}}\right).
\end{equation}

The down-spin density $n_\downarrow$ can be explicitly written in
terms of the external magnetic field $H$ by evaluating
Eq.~\eqref{equiv1}. Using the property that $Z_{sc}(k)\approx
Z_{sc}(k_0)+O(1/c^2)$ for $c\gg 1$ and $k_0\approx \pi n_c/(1+2\ln
2/\gamma)$ (see Eq.~\eqref{fermi-p}), we find that
\begin{equation}
n_{\downarrow}=\frac{n_c}{2}\left[1-\frac{4}{\pi^{2}}\left(\frac{H}{H_{c}}\right)\right].
\label{downspin}
\end{equation}
By substituting the expression \eqref{downspin} into
Eqs.~\eqref{zss} and \eqref{zsc}, the dressed charges in the strong
coupling regime $c\gg 1$ and a weak magnetic field $H\ll 1$ are
explicitly determined in terms of the fixed particle density $n_c$
and the external magnetic field $H$.

\subsection{The limit $H\rightarrow H_{c}$ for $\gamma\gg 1/\sqrt{1-H/H_c}$}
When $H\geq H_{c}$, the ground state of the system is ferromagnetic.
Correspondingly the Fermi point $\lambda_0$ becomes zero. Before
solving the dressed charged matrix for $H$ approaching the critical
field $H_{c}$ from below, we have to know how $\lambda_{0}$ behaves
in this vicinity. The spin part of the TBA equation  at $T=0$ (i.e.,
$\phi_1^{(0)}(\lambda)$ in Eq.~\eqref{phi-0} or equivalently the
dressed energy $\varepsilon_s(\lambda)$ in Eq.~\eqref{dress}) is
approximately
\begin{align}
\nonumber \varepsilon_s(\lambda) &=
H+\int_{-k_{0}}^{k_{0}}a_{1}(\lambda-k)
\varepsilon_c(k)dk-\int_{-\lambda_{0}}^{\lambda_{0}}a_{2}(\lambda-\mu)\varepsilon_s(\mu)d\mu
\\ &\approx
H-2\pi P_0 a_1(\lambda)-2\lambda_0 a_2(\lambda) \varepsilon_s(0)
\label{Phi1}
\end{align}
which is derived by approximating the first integral as in
\eqref{appro} and expanding the second integral around
$\lambda_0\approx 0$.  Let us assume $\gamma\gg 1/\sqrt{1-H/H_c}\gg
1$. In this region, one sees from Eq.~\eqref{Hc} that
Eq.~\eqref{Phi1} further reduces to
\begin{equation}
\varepsilon_s(\lambda)=H-\frac{H_c}{1+4\lambda^2/c^2}-\frac{2\lambda_0}{\pi c}
\frac{\varepsilon_s(0)}{1+\lambda^2/c^2}.
\end{equation}
With this result, we can explicitly find
$\varepsilon_s(0)$ to be
\begin{equation}
\varepsilon_s(0)=-\frac{H_{c}-H}{1+2\lambda_{0}/\pi c}.
\end{equation}

The last step to derive an explicit expression for $\lambda_{0}$ is
to use the condition $\varepsilon_{s}(\pm\lambda_{0})=0$. This
gives
\begin{equation}
H-\frac{H_{c}}{1+4\lambda_{0}^{2}/c^{2}}+\frac{2\lambda_{0}}{\pi
c}\frac{H_{c}-H}{(1+\lambda_{0}^{2}/c^{2})(1+2\lambda_{0}/\pi c)}=0.
\end{equation}
Multiplying the equation with each denominator and then ignoring the
terms of order $O(\lambda_{0}^{3}/c^{3})$ or higher yields the
equation
\begin{equation}
H-H_{c}+\frac{5\lambda_{0}^{2}}{c^{2}}H-\frac{\lambda_{0}^{2}}{c^{2}}H_{c}=0.
\end{equation}
After rearranging the terms and using the fact that $H\rightarrow
H_{c}$, we arrive at the result
\begin{equation}
\lambda_{0}\approx \frac{c}{2}\sqrt{1-\frac{H}{H_{c}}}
\label{lambda}
\end{equation}
which is also similar to the result obtained for the 1D Hubbard model \cite{Frahm1991} and
the XXZ Heisenberg chain \cite{Bogoliubov1986}.

With this expression, we can evaluate the dressed charge matrix
explicitly in terms of $H$. For $Z_{cs}(\lambda_0)$, from \eqref{D2}
\begin{equation}
Z_{cs}(\lambda) \approx 2\pi a_1(\lambda) n_c-2\lambda_0
a_2(\lambda)Z_{cs}(0).
\end{equation}
Here an approximation similar to the above and the property
\eqref{equiv2} have been used. Solving for $Z_{cs}(0)$ and neglecting
terms of order $O(1/c^2)$, we have $Z_{cs}(0)\approx 4/\gamma$.
Substituting this together with \eqref{lambda} implies that
\begin{equation}
Z_{cs}(\lambda_{0})\approx
\frac{4}{\gamma}\left(1-\frac{1}{\pi}\sqrt{1-\frac{H}{H_{c}}}\right).
\end{equation}
$Z_{cc}(k_0)$ is easily obtained by substituting the above results
into the approximation form $Z_{cc}(k_0)\approx 1+2\lambda_0 a_1(k_0) Z_{cs}(0)$
of Eq.~\eqref{D1}. The result reads
\begin{equation}
Z_{cc}(k_0)\approx 1+\frac{8}{\pi\gamma}\sqrt{1-\frac{H}{H_{c}}}.
\end{equation}

The remaining dressed charges $Z_{ss}(\lambda_0)$ and $Z_{sc}(k_0)$
are also evaluated by similar calculations. For
$Z_{ss}(\lambda_0)$, Eq.~\eqref{D4} becomes $Z_{ss}(\lambda)=1+2\pi
n_{\downarrow}a_1(\lambda)-2\lambda_0 a_2(\lambda) Z_{ss}(0)$, and
hence $Z_{ss}(\lambda_0)=1+4 n_{\downarrow}/c-2\lambda_0/(\pi c)$.
The density of down-spin fermions $n_{\downarrow}$ is evaluated by
applying the same method as in the above to Eq.~\eqref{density}, with result
\begin{equation}
n_{\downarrow} =\int_{-\lambda_0}^{\lambda_0} \rho_s(\lambda)
d\lambda \approx 2\lambda_0
\rho_s(0)\approx\frac{2n_c}{\pi}\sqrt{1-\frac{H}{H_c}}.
\end{equation}
This in turn gives
\begin{equation}
Z_{ss}(\lambda_0)\approx 1-\frac{1}{\pi}\sqrt{1-\frac{H}{H_c}}+
\frac{8}{\pi \gamma}\sqrt{1-\frac{H}{H_c}}.
\end{equation}
Likewise,
\begin{equation}
Z_{sc}(k_0)\approx\frac{2}{\pi}\sqrt{1-\frac{H}{H_c}}.
\end{equation}

\begin{figure}
\centering
\begin{tabular}{cc}
\epsfig{file=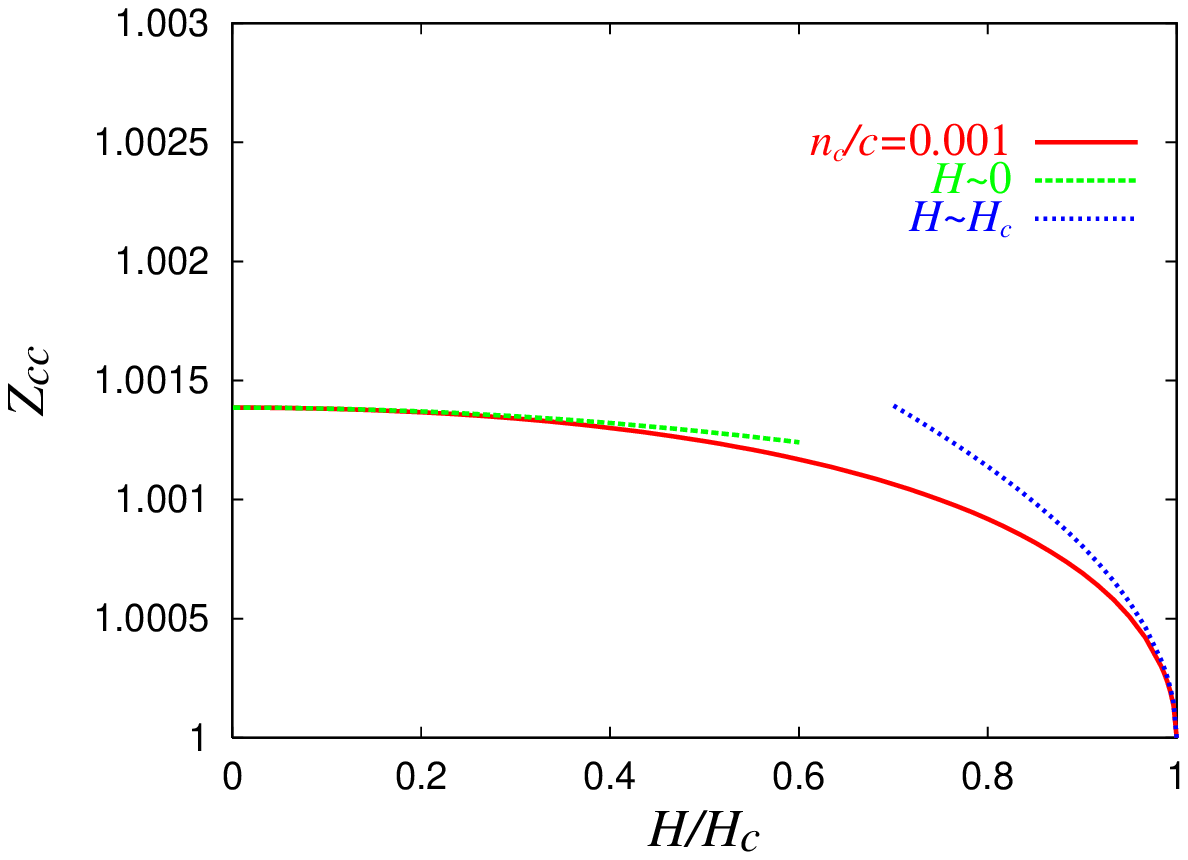,width=0.5\linewidth,clip=} &
\epsfig{file=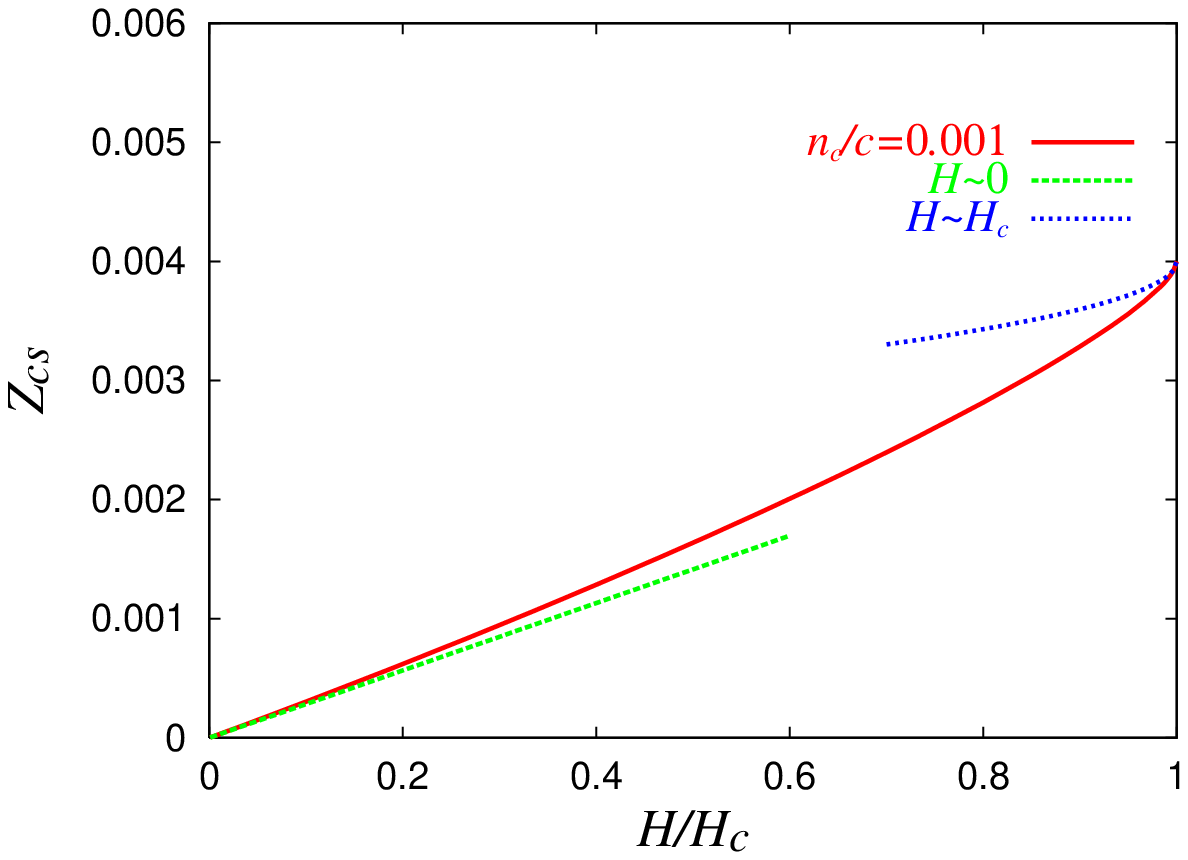,width=0.5\linewidth,clip=} \\
\epsfig{file=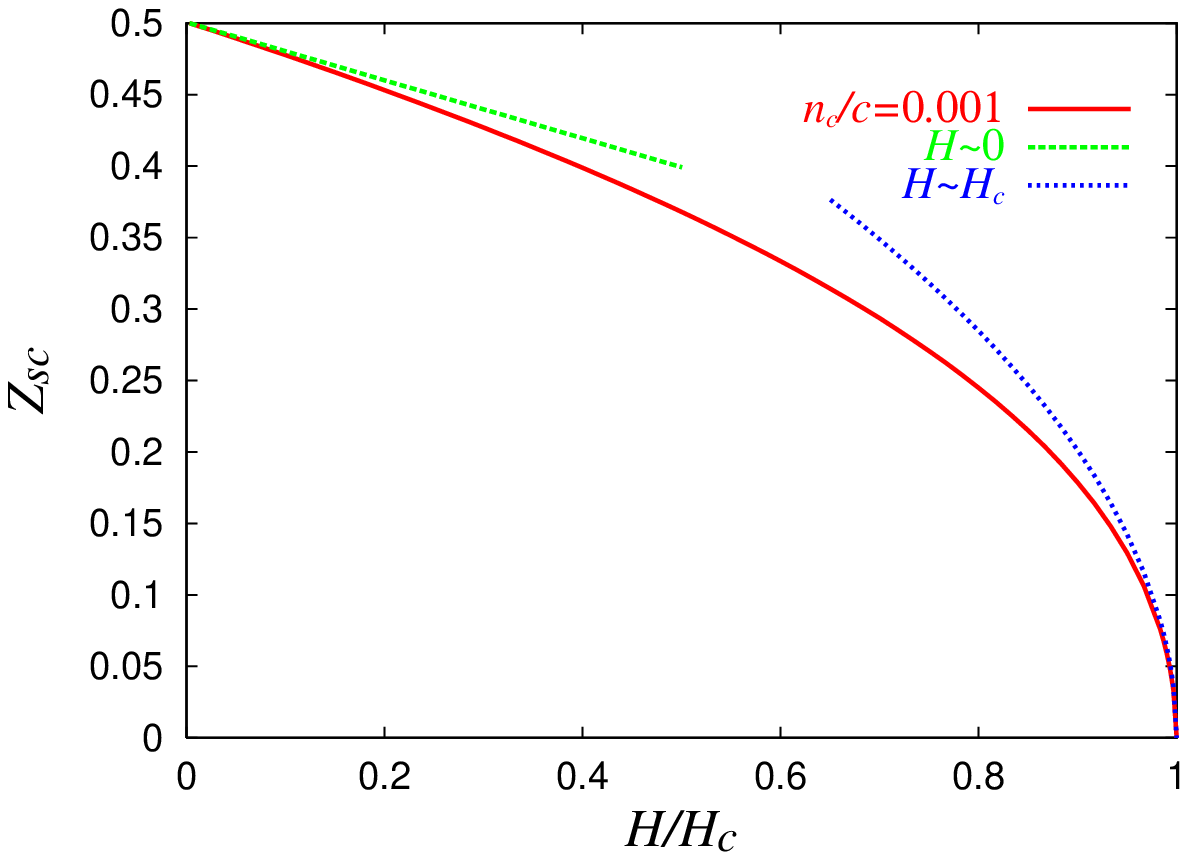,width=0.5\linewidth,clip=} &
\epsfig{file=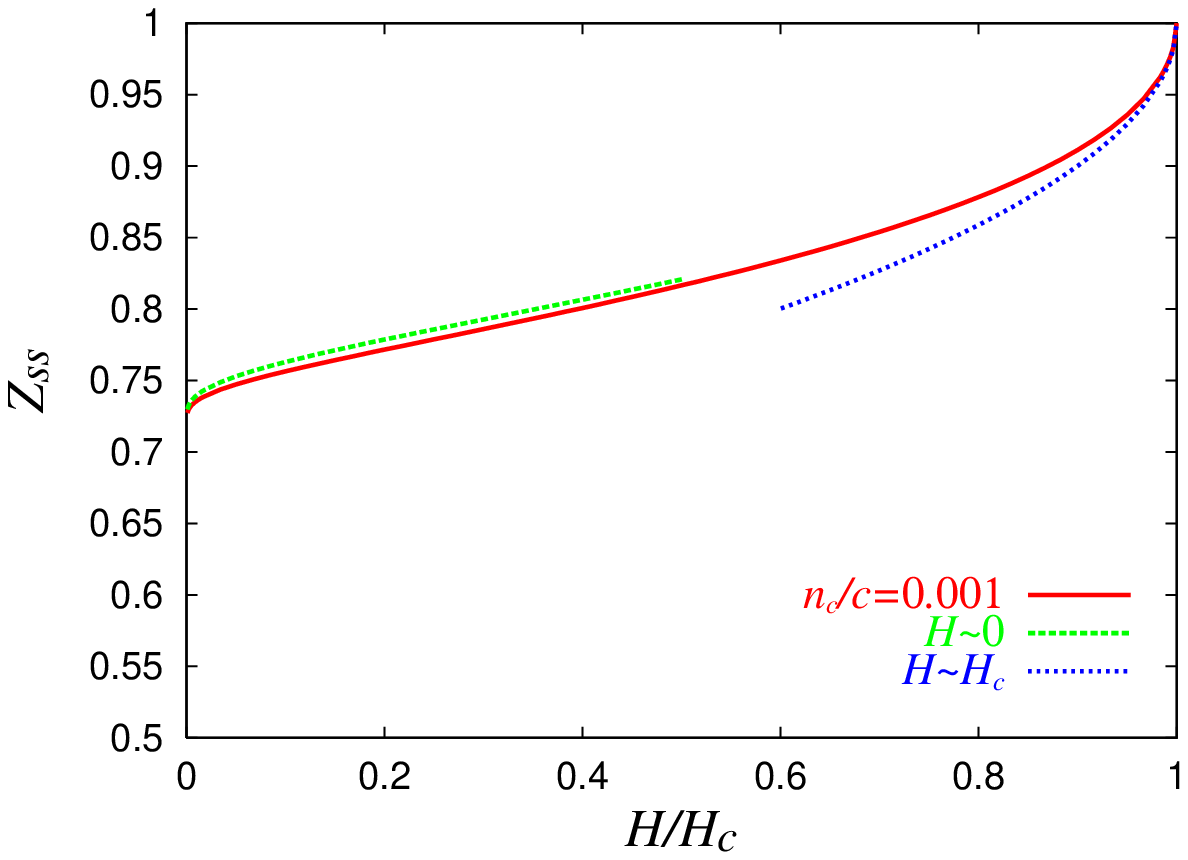,width=0.5\linewidth,clip=}
\end{tabular}\caption{(Color online) This figure shows
a comparison between the numerical solutions (solid lines) and the leading order
corrections to the dressed charges $Z_{cc}(k_{0})$,
$Z_{cs}(\lambda_{0})$, $Z_{sc}(k_{0})$ and $Z_{ss}(\lambda_{0})$ in
the limits $H\ll 1$ and $H\rightarrow H_{c}$ for
$n_{c}/c=0.001$.}\label{fig:Z-lim}
\end{figure}

In FIG.~\ref{fig:Z-lim}, we compare the numerical solutions to the
leading order solutions for the dressed charges in both limits $H\ll
1$ and $H\rightarrow H_{c}$. It shows good agreement between our
leading order solutions and the numerical solutions at points not
far from $H=0$ and $H=H_{c}$.

\section{Correlation functions}
\label{section-corr}

We turn now to the calculation of the long distance asymptotics of 
various correlation functions and scaling dimensions 
which have been obtained for the 1D Hubbard model in terms of the dressed charge 
formalism \cite{book3,Frahm1990,Frahm1991}. Our results extend these results into the strong but finite 
coupling regime for the spin-1/2 repulsive fermion model. 

The spin-1/2 repulsive fermion model is gapless and thus critical at
zero temperature. At $T=0$, the correlation functions decay as some
power  of distance governed by the critical exponent
which we shall denote conventionally by $\theta$. For $T>0$
the decay is exponential. It was shown that conformal invariance
leads to universality classes of critical theories that are related to
the central charge $C$ related to
the underlying Virasoro algebra \cite{Belavin1984}. The critical
behavior of the model under consideration is described by the direct product of two
Virasoro algebras, one characterizing the spin degree and the other
characterizing the charge degree. Both Virasoro algebras have
central charge $C=1$.

The general two-point correlation function for primary fields
$\varphi$ with conformal dimensions $\Delta^{\pm}_{c,s}$ at $T=0$
and $T>0$ are given by
\begin{equation}
\langle\varphi(x,t)\varphi(0,0)\rangle=\frac{\exp(-2iD_{c}(k_{F\uparrow}+k_{F\downarrow})x)\exp(-2iD_{s}k_{F\downarrow}x)}
{(x-iv_{c}t)^{2\Delta^{+}_{c}}(x+iv_{c}t)^{2\Delta^{-}_{c}}(x-iv_{s}t)^{2\Delta^{+}_{s}}(x+iv_{s}t)^{2\Delta^{-}_{s}}}
\end{equation}
and
\begin{eqnarray}
\nonumber\lefteqn{\langle\varphi(x,t)\varphi(0,0)\rangle_{T}} \\ &&
\nonumber
=\exp(-2iD_{c}(k_{F\uparrow}+k_{F\downarrow})x)\exp(-2iD_{s}k_{F\downarrow}x)
\\ && \nonumber\times\left(\frac{\pi T}{v_{c}\sinh(\pi
T(x-iv_{c}t)/v_{c})}\right)^{2\Delta_{c}^{+}} \left(\frac{\pi
T}{v_{c}\sinh(\pi T(x+iv_{c}t)/v_{c})}\right)^{2\Delta_{c}^{-}}
\\ && \times\left(\frac{\pi T}{v_{s}\sinh(\pi
T(x-iv_{s}t)/v_{s})}\right)^{2\Delta_{s}^{+}}\left(\frac{\pi
T}{v_{s}\sinh(\pi T(x+iv_{s}t)/v_{s})}\right)^{2\Delta_{s}^{-}}
\end{eqnarray}
where $k_{F\downarrow,\uparrow}$ are the Fermi momenta, $0<x\leq L$
and $-\infty<t<\infty$ is Euclidean time. The conformal dimensions
of the fields can be written in terms of the elements of the dressed
charge matrix as
\begin{equation}
2\Delta_{c}^{\pm}(\Delta
N_{c,s},N^{\pm}_{c,s},D_{c,s})=\left(Z_{cc}D_{c}+Z_{sc}D_{s}\pm
\frac{Z_{ss}\Delta N_{c}-Z_{cs}\Delta N_{s}}{2\det
Z}\right)^{2}+2N_{c}^{\pm}
\end{equation}
and
\begin{equation}
2\Delta_{s}^{\pm}(\Delta
N_{c,s},N^{\pm}_{c,s},D_{c,s})=\left(Z_{cs}D_{c}+Z_{ss}D_{s}\pm
\frac{Z_{cc}\Delta N_{s}-Z_{sc}\Delta N_{c}}{2\det
Z}\right)^{2}+2N_{s}^{\pm}.
\end{equation}

The non-negative integers $\Delta N_{\alpha}$, $N_{\alpha}^{\pm}$
and the parameter $D_{\alpha}$ where $\alpha=c,s$ represent the
three types of low-lying excitations. Here $\Delta N_{\alpha}$
denotes the change in the number of down-spin fermions.
$N_{\alpha}^{\pm}$ characterizes particle-hole excitations where
$N_{\alpha}^{+}$ ($N_{\alpha}^{-}$) is the number of occupancies
that a particle at the right (left) Fermi level jumps to.
$N_{\alpha}^{\pm}$ also enumerates the descendent fields for the
primary fields $\varphi$. And lastly, $D_{\alpha}$ represents
fermions that are backscattered from one Fermi point to the other.
They are restricted by the condition
\begin{equation}
D_{c}\equiv\frac{\Delta N_{s}+\Delta N_{s}}{2}\quad (\mathrm{mod}1),
\qquad D_{s}\equiv\frac{\Delta N_{c}}{2}\quad (\mathrm{mod}1).
\label{mod}
\end{equation}

We want to find the asymptotic behavior of the general two-point
correlation functions for the operators $O(x,t)$, namely $\langle
O(x,t)O^{\dagger}(0,0)\rangle$. The operators can be written as a
linear combination of primary fields with conformal dimensions
$\Delta^{\pm}_{c,s}$ and their descendent fields. Noting that the
correlation functions for fields with different conformal dimensions
are zero, we can express the correlation functions at $T=0$ and
$T>0$ respectively as
\begin{equation}
\langle O(x,t)O^{\dagger}(0,0)\rangle=\sum_{\mathbf{n}} A(\Delta
N_{c,s},N^{\pm}_{c,s},D_{c,s})\frac{\exp(-2iD_{c}(k_{F\uparrow}+k_{F\downarrow})x)
\exp(-2iD_{s}k_{F\downarrow}x)}{(x-iv_{c}t)^{2\Delta^{+}_{c}}(x+iv_{c}t)^{2\Delta^{-}_{c}}
(x-iv_{s}t)^{2\Delta^{+}_{s}}(x+iv_{s}t)^{2\Delta^{-}_{s}}}
\end{equation}
and
\begin{eqnarray}
\nonumber\lefteqn{\langle O(x,t)O^{\dagger}(0,0)\rangle_{T}} \\ &&
\nonumber =\sum_{\mathbf{n}} A(\Delta N_{c,s},N^{\pm}_{c,s},D_{c,s})
\exp(-2iD_{c}(k_{F\uparrow}+k_{F\downarrow})x)\exp(-2iD_{s}k_{F\downarrow}x)
\\ && \nonumber\times\left(\frac{\pi T}{v_{c}\sinh(\pi
T(x-iv_{c}t)/v_{c})}\right)^{2\Delta_{c}^{+}} \left(\frac{\pi
T}{v_{c}\sinh(\pi T(x+iv_{c}t)/v_{c})}\right)^{2\Delta_{c}^{-}}
\\ && \times\left(\frac{\pi T}{v_{s}\sinh(\pi
T(x-iv_{s}t)/v_{s})}\right)^{2\Delta_{s}^{+}}\left(\frac{\pi
T}{v_{s}\sinh(\pi T(x+iv_{s}t)/v_{s})}\right)^{2\Delta_{s}^{-}}
\end{eqnarray}
where $\mathbf{n}$ denotes the set of quantum numbers
\begin{equation}
\mathbf{n}=(\Delta N_{c,s},N^{\pm}_{c,s},D_{c,s})
\end{equation}
which are determined by the condition given in (\ref{mod}) and the
selection rules for the form factors while performing a spectral
decomposition of the correlation functions \cite{Cardy1986}.

Let us consider the correlation functions of operators which are
written in terms of the field operators $\psi_{\sigma}(x,t)$ where
$\sigma=\uparrow,\downarrow$. They obey the canonical commutation
relations
\begin{eqnarray}
\nonumber \{\psi_{\sigma}(x,t),\psi_{\sigma'}^{\dagger}(x,t')\} &=&
\delta_{\sigma\sigma'}\delta(x-x')
\\ \{\psi_{\sigma}(x,t),\psi_{\sigma}(x,t')\} &=&
\{\psi_{\sigma}^{\dagger}(x,t),\psi_{\sigma}^{\dagger}(x,t')\}=0.
\end{eqnarray}
Here we consider the following correlation functions:

(i) One particle Green's function:
\begin{equation}
G_{\sigma}(x,t)=\langle\psi_{\sigma}(x,t)\psi_{\sigma}^{\dagger}(0,0)\rangle.
\end{equation}

(ii) Charge density correlation function:
\begin{equation}
G_{nn}(x,t)=\langle n(x,t)n(0,0)\rangle
\end{equation}
where
\begin{equation}
n(x,t)=n_{\uparrow}(x,t)+n_{\downarrow}(x,t),\qquad
n_{\sigma}(x,t)=\psi_{\sigma}^{\dagger}(x,t)\psi_{\sigma}(x,t).
\end{equation}

(iii) Longitudinal spin-spin correlation function:
\begin{equation}
G^{z}(x,t)=\langle S^{z}(x,t)S^{z}(0,0)\rangle
\end{equation}
where
\begin{equation}
S^{z}(x,t)=\frac{1}{2}(n_{\uparrow}(x,t)-n_{\downarrow}(x,t)).
\end{equation}

(iv) Transverse spin-spin correlation function:
\begin{equation}
G^{\perp}(x,t)=\langle S^{+}(x,t)S^{-}(0,0)\rangle
\end{equation}
where
\begin{equation}
S^{+}(x,t)=\psi_{\uparrow}^{\dagger}(x,t)\psi_{\downarrow}(x,t),\qquad
S^{-}(x,t)=\psi_{\downarrow}^{\dagger}(x,t)\psi_{\uparrow}(x,t).
\end{equation}

(v) Pair correlation function:
\begin{equation}
G_{p}(x,t)=\langle\psi_{\downarrow}(x,t)\psi_{\uparrow}(x,t)
\psi_{\uparrow}^{\dagger}(0,0)\psi_{\downarrow}^{\dagger}(0,0)\rangle.
\end{equation}

For each of the correlation functions considered above, the values of
$\mathbf{n}$ are given by
\begin{eqnarray}
\nonumber G_{\uparrow}(x,t) &:& (\Delta N_{c}=1,\Delta
N_{s}=0,D_{c}\in\mathbb{Z}+1/2,D_{s}\in\mathbb{Z}+1/2) \\
\nonumber G_{\downarrow}(x,t) &:& (\Delta N_{c}=1,\Delta
N_{s}=1,D_{c}\in\mathbb{Z},D_{s}\in\mathbb{Z}+1/2) \\ \nonumber
G_{nn}(x,t) &:& (\Delta N_{c}=0,\Delta
N_{s}=0,D_{c}\in\mathbb{Z},D_{s}\in\mathbb{Z}) \\ \nonumber
G^{z}(x,t) &:& (\Delta N_{c}=0,\Delta
N_{s}=0,D_{c}\in\mathbb{Z},D_{s}\in\mathbb{Z}) \\ \nonumber
G^{\perp}(x,t) &:& (\Delta N_{c}=0,\Delta
N_{s}=1,D_{c}\in\mathbb{Z}+1/2,D_{s}\in\mathbb{Z}) \\ \nonumber
G_{p}(x,t) &:& (\Delta N_{c}=2,\Delta
N_{s}=1,D_{c}\in\mathbb{Z}+1/2,D_{s}\in\mathbb{Z}).
\end{eqnarray}
with $N^{\pm}_{c,s}\in\mathbb{Z}_{\geq 0}$ for every case. The explicit results for  these correlation functions for $H\ll 1$ and $H\to H_c$ are given   in Appendices  \ref{CF-1} and \ref{CF-2}, which include   the order of $1/\gamma$ corrections in the critical exponents.

\section{Conclusion}\label{section-conclude}

We have derived the low temperature thermodynamics and long distance
asymptotics of correlation functions for the spin-1/2 repulsive
delta-function interacting Fermi gas with an external field by means
of the thermodynamic Bethe ansatz method and dressed charge
formalism. With the help of Wiener-Hopf techniques we have
calculated the low temperature free energy and thermodynamics and
found that the low energy physics can be described by a spin-charge
separated theory of a Tomonaga-Luttinger liquid and an
antiferromagnetic spin Heisenberg chain. The dressed charge
equations have been solved analytically for a small external field
$H\to 0$ and a large external field $H\to H_c$ using the Wiener-Hopf
method. We have also calculated the conformal dimensions for many
correlation functions including the one particle Green's function,
the charge density correlation function and pairing correlation, as
given in Section \ref{section-corr}.

In particular, the explicit form of the critical exponents which we
have obtained in terms of the external magnetic field and the
interaction strength up to $1/c$ corrections extends the
known results obtained for the 1D Hubbard model in the infinite coupling limit 
\cite{book3,Frahm1990}. 
They provide insight into
understanding the critical behaviour of interacting fermions in 1D.
The result for the free energy at low temperature shows a universal
signature of Tomonaga-Luttinger liquids where the leading low
temperature contributions are solely dependent on the spin and
charge velocities. It is to be expected that this universal nature
can be tested via the finite temperature density profiles of the
repulsive Fermi gas in an harmonic trap. This opens a way to
experimentally observe how the low temperature thermodynamics of a
1D many-body system naturally separates into two free Gaussian field
theories.

We have also presented results for the low temperature
thermodynamics which extend beyond the range covered by spin-charge separation theory.
As effective as it is, the Wiener-Hopf method does not allow the
full derivation of the equation of state for temperatures beyond the
Tomonaga-Luttinger liquid regime. 
This is because the Tomonaga-Luttinger low temperature physics does not 
contain enough information on thermal fluctuations necessary to 
describe the quantum critical regime. 
This restricts access to quantum criticality in the whole physical regime.
However, as we have demonstrated, the polylog formalism   is suitable for 
the study of  low temperature and strong coupling in the $\mu-H$
phase plane for weak magnetic field.  The pressure  we obtained captures    
the essential spin density and charge density fluctuations at criticality and  may  
possibly be used to test the spin-charge separation theory  in experiment with ultracold atomic fermions.  
This points to the further study of quantum
criticality in 1D interacting Fermi gases with repulsive
interaction, as has been done recently for attractive interaction and for the  Lieb-Liniger Bose gas 
\cite{GHo,GBat}.

This work has been supported by the Australian Research Council. The
authors thank Professor Tin-Lun Ho and  Professor Rafael I.
Nepomechie for helpful discussions.

\clearpage

\appendix
\section{The ground state properties}\label{ground}

Here we briefly describe the ground state properties of spin-1/2
fermions with repulsive interaction. The full spectrum of the
Hamiltonian can be obtained by the BA method \cite{Yang, Gaudin}. In
the thermodynamic limit, the ground state properties are
characterized by two Fermi seas made up of charges and (down) spins.
The distribution functions $\rho_c(k)$ of charges with holon
momentum $k$, and  $\rho_s(k)$ of down spins with spinon rapidity
$\lambda$ are written as integral equations,
\begin{align}
\rho_c(k)&=\frac{1}{2\pi}+\int_{-\lambda_0}^{\lambda_0}
a_1(k-\lambda)\rho_s(\lambda) d\lambda
\nonumber \\
\rho_s(\lambda)&=\int_{-k_0}^{k_0}a_1(\lambda-k)\rho_c(k)
dk-\int_{-\lambda_0}^{\lambda_0} a_2(\lambda-\mu)\rho_s(\mu)d\mu
\label{density}
\end{align}
where the function $a_n(x)$ is given by \eqref{kernel} and $\pm k_0$
and $\pm \lambda_0$ correspond to the Fermi  points. The density of
the fermions $n_c=n_{\uparrow}+n_{\downarrow}$ ($n_\sigma$ denotes
the density of spin-$\sigma$ fermions) and the density of the
down-spin fermions $n_{\downarrow}$ are respectively given by
\begin{equation}
\int_{-k_0}^{k_0} \rho_c(k) dk=n_c,\qquad
\int_{-\lambda_0}^{\lambda_0} \rho_s(\lambda)
d\lambda=n_{\downarrow}. \label{density2}
\end{equation}

The ground state energy per unit length (denoted by $E_0$) is
\begin{equation}
E_0-\mu n_c=-P_0=\int_{-k_0}^{k_0} (k^2-\mu-H/2)\rho_c(k)d k+ H
\int_{-\lambda_0}^{\lambda_0}\rho_s(\lambda) d\lambda
\end{equation}
where $P_0$ is the pressure at zero-temperature (see
\eqref{Pressure} for finite temperature). The ground state
properties are also described in terms of the charge dressed energy
$\varepsilon_c(k)$ and the spin dressed energy
$\varepsilon_s(\lambda)$ as
\begin{align}
\varepsilon_c(k)&=k^2-\mu-H/2+\int_{-\lambda_0}^{\lambda_0}
a_1(k-\lambda)\varepsilon_s(\lambda) d\lambda \nonumber \\
\varepsilon_s(\lambda)&=H+\int_{-k_0}^{k_0}a_1(\lambda-k)\varepsilon_c(k) dk-
\int_{-\lambda_0}^{\lambda_0}a_2(\lambda-\mu)\varepsilon_s(\mu) d\mu.
\label{dress}
\end{align}

The above dressed energies define the energy bands. The ground state
corresponds to the fillings of $\varepsilon_c(k)\le 0$ and
$\varepsilon_s(\lambda)\le 0$. Thus the Fermi points $\pm k_0$ and
$\pm \lambda_0$ are given by the conditions
\begin{equation}
\varepsilon_c(\pm k_0)=0, \qquad \varepsilon_s(\pm \lambda_0)=0.
\label{condition}
\end{equation}
Using the above dressed energies,  $E_0$ is written as
\begin{equation}
E_0-\mu n_c=-P_0=\frac{1}{2\pi}\int_{-k_0}^{k_0}\varepsilon_c(k) dk.
\label{groundstate}
\end{equation}
One can immediately realize that the dressed energies
$\varepsilon_c(k)$ and $\varepsilon_s(\lambda)$ respectively
correspond to $\varepsilon(k)$ and $\phi_1(\lambda)$ in the TBA
equations \eqref{TBA-k} and \eqref{TBA-lambda} in the limit $T\to
0$.

Let us evaluate the value of the critical field $H=H_c$. At this
point, the density of down-spin electrons is zero
($n_{\downarrow}=0$) i.e., $\lambda_0=0$. Therefore the expressions
\eqref{density} and \eqref{dress} are significantly simplified.
Inserting $\rho_c(k)=1/(2\pi)$ into \eqref{density2}, one has
$k_0=\pi n_c$. The Fermi point $k_0$ is also calculated by
$\varepsilon_c(k)=k^2-\mu-H_c/2$ \eqref{dress} and the condition
\eqref{condition}, i.e. $k_0^2=\mu+H_c/2$. Substituting these
expressions into $\varepsilon_s(\lambda)$ and using the condition
\eqref{condition}, one finally arrives at
\begin{equation}
H_c=\left(\frac{c^2}{2 \pi }+2 \pi  n_c^2\right) \tan
^{-1}\left(\frac{2 \pi n_c}{c}\right)-c n_c.
\end{equation}
The pressure $P_0$ at $H=H_c$ is
\begin{equation}
P_0=\frac{2}{3}\pi^2 n_c^3.
\end{equation}
In the strong coupling limit $c\gg 1$, $H_c$ and $P_0$ are given by
\begin{equation}
H_c\approx\frac{8\pi^2 n_c^2 }{3\gamma}\left(1-\frac{4 \pi^2}{5
\gamma^2}\right),\qquad P_0\approx\frac{n_c \gamma
H_c}{4}\left(1+\frac{4 \pi^2}{5 \gamma^2}\right). \label{Hc}
\end{equation}

\section{The Wiener-Hopf method}\label{WH}

In this appendix, we briefly review how to solve the integral
equations appearing in the main text (see Eq.~\eqref{scaling} for
instance) by using the Wiener-Hopf method.

Consider a Wiener-Hopf integral equation
\begin{equation}
y(\lambda)=g(\lambda)+\int_0^\infty K(\lambda-\mu)y(\mu) d\mu
\label{WH-eq}
\end{equation}
which determines an unknown function $y(\lambda)$ where $K(\lambda)$
is defined by Eq.~\eqref{kernel}, and the driving term $g(\lambda)$
defined on the entire real axis is assumed to be a known function.
By Fourier transforming Eq.~\eqref{WH-eq}, we obtain
\begin{equation}
\widehat{y}_+(\omega)=\left(1-\widehat{K}(\omega)\right)^{-1}
\left(\widehat{g}(\omega)-\widehat{y}_-(\omega)\right) \label{F1}
\end{equation}
where the functions $\widehat{y}_{\pm}(\omega)$ are defined by
\begin{equation}
\widehat{y}_{\pm}(\omega)=\int_{-\infty}^{\infty} \theta(\pm
\lambda) y(\lambda)e^{i\lambda\omega} d\lambda \label{decomp-y}
\end{equation}
with $\theta(\lambda)$ representing the Heaviside step function. The
function $\widehat{K}(\omega)$ can be explicitly written as
\begin{equation}
\widehat{K}(\omega)=\frac{e^{-c|\omega|/2}}{2\cosh(c|\omega|/2)}.
\end{equation}
Eqs.~\eqref{decomp-y} denote a decomposition of
$\widehat{y}(\omega)$ into the sum of functions analytic on the
upper and lower half-planes, respectively, with $\widehat{y}(\omega)
=\widehat{y}_+(\omega)+\widehat{y}_-(\omega)$. Hereafter we assume
that $\widehat{y}_\pm (\infty)=\widehat{g}(\infty)=0$.

To solve the equation, we factorize the term $1-\widehat{K}(\omega)$
as
\begin{equation}
\left(1-\widehat{K}(\omega)\right)^{-1}=G_+(\omega)G_-(\omega),
\qquad \lim_{\omega \to \infty}G_\pm (\omega)=1
\label{factorization}
\end{equation}
where $G_+(\omega)$ ($G_-(\omega)$) is a function which is analytic
and nonzero in the upper (lower) half-plane. Since
$\widehat{K}(\omega)=\widehat{K}(-\omega)$, one finds
$G_+(\omega)/G_-(-\omega)=G_+(-\omega)/G_-(\omega)$. Therefore
$G_+(\omega)/G_-(-\omega)$ is a bounded entire function. Liouville's
theorem and the asymptotics of $G_\pm(\omega)$ yield
\begin{equation}
G_+(\omega)=G_-(-\omega).
\end{equation}

Using the factorization equation \eqref{factorization},
Eq.~\eqref{F1} becomes
\begin{equation}
\frac{\widehat{y}_+(\omega)}{G_+(\omega)}-\Phi_+(\omega)
=\Phi_-(\omega)-G_-(\omega)\widehat{y}_-(\omega) \label{decomp-A1}
\end{equation}
where $\Phi_\pm(\omega)$ are functions assumed to be
analytic and bounded in the upper and lower half planes, respectively,
\begin{equation}
G_-(\omega)\widehat{g}(\omega)=\Phi_+(\omega)+\Phi_-(\omega).
\label{decomp-A2}
\end{equation}

From Eq.~\eqref{decomp-A1}, the function
$\widehat{y}_+(\omega)/G_+(\omega)-\Phi_+(\omega)$ is a bounded
entire function, and hence is a constant according to Liouville's
theorem. Considering the asymptotics, we obtain
\begin{equation}
\widehat{y}_+(\omega)=G_+(\omega)\left(\Phi_+(\omega)-\Phi_+(
\infty)\right), \qquad
\widehat{y}_-(\omega)=\frac1{G_-(\omega)}\left(\Phi_-(\omega)-\Phi_-(\infty)\right).
\label{decomp-Phi}
\end{equation}
A formula useful for  practical calculations is
\begin{equation}
y(0)=\frac{1}{2\pi}\int_{-\infty}^\infty \widehat{y}(\omega) e^{-i
\omega \epsilon} d\omega =\frac{1}{2\pi}\int_{C^{-}}
\widehat{y}_+(\omega) e^{-i \omega \epsilon} d\omega =-i
\lim_{\omega\to\infty} \omega \widehat{y}_+(\omega)
\label{formula-y0}
\end{equation}
where $C^{-}$ is a semi-circular path on the lower half-plane. We
have used the fact that the sum of residues in the lower half-plane
is equal to the residue at the point at infinity.

Next we will determine the explicit forms of
$\widehat{y}_{\pm}(\omega)$ by adopting some specific driving terms
$g(\lambda)$ appearing in the main text. First let us  determine the
factors $G_\pm(\omega)$. Using the explicit form of
$\widehat{K}(\omega)$ in \eqref{kernel}, we have
\begin{equation}
2e^{-c|\omega|/2}\cosh(c\omega/2)=G_+(\omega)G_-(\omega).
\end{equation}
The well-known relation $\Gamma(1/2+x)\Gamma(1/2-x)=\pi/\cos\pi x$
\cite{Gradshteyn} together with the asymptotic form
$\Gamma(x)\approx\sqrt{2\pi}x^{x-\frac{1}{2}}e^{-x}$ for $|x|\gg 1$
and the condition for $G_{\pm}(\omega)$ in Eq.~\eqref{factorization}
yield
\begin{equation}
G_+(\omega)=G_-(-\omega)=\frac{\sqrt{2\pi}}{\Gamma(\frac{1}{2}-\frac{i
c \omega}{2\pi})} \left(\frac{2\pi e}{\epsilon-i c
w}\right)^{\frac{i c \omega}{2\pi}} \label{g-func0}
\end{equation}
where $\epsilon\to 0+$. Useful special values are
\begin{equation}
G_\pm(0)=\sqrt{2}, \qquad G_+\left(\frac{\pi
i}{c}\right)=G_-\left(-\frac{\pi i}{c}\right)=\sqrt{\frac{\pi}{e}}.
\label{g-func}
\end{equation}

\subsection{The case $g(\lambda)=a+b s(\lambda+\lambda_0)$}
Set the driving term to be $g(\lambda)=a+b s(\lambda+\lambda_0)$,
where $a,b\in \mathbb{C}$. Taking the Fourier transform yields
\begin{equation}
\widehat{g}(\omega)=2\pi a \delta(\omega)+
\frac{b e^{-i \lambda_0 \omega}}{2\cosh(c\omega/2)}.
\label{Fourier}
\end{equation}
Let us decompose the function as in \eqref{decomp-A2}. The first term can be easily
decomposed by using
\begin{equation}
\delta(\omega)=\frac{1}{2\pi i}\left(\frac{1}{\omega-i \epsilon}-
\frac{1}{\omega+i\epsilon}\right) \qquad (\epsilon\to +0).
\end{equation}

On the other hand, the second term in \eqref{Fourier} is a
meromorphic function whose poles are simple poles (denoted by
$\omega_n$)  located at
\begin{equation}
\omega_n=\frac{\pi i}{c}(2n+1) \qquad (n\in\mathbb{Z}). \label{pole}
\end{equation}
Thus the decomposition of the function $1/\cosh (c\omega/2)$ reads
\begin{eqnarray}
\frac{1}{\cosh(c\omega/2)} &=& Q_+(\omega)+Q_-(\omega), \nonumber \\
Q_+(\omega) &=& \frac{2i}{c}\sum_{n=0}^\infty
\frac{(-1)^{n}}{\omega+\omega_n}, \nonumber \\
Q_-(\omega)&=& \frac{1}{\cosh(c\omega/2)}-
\frac{2i}{c}\sum_{n=0}^\infty \frac{(-1)^{n}}{\omega+\omega_n}.
\end{eqnarray}
Using this, we can express the function $f_-(\omega)/\cosh(c
\omega/2)$, where $f_-(\omega)$ is any function which is analytic
and bounded in the lower half-plane, as
\begin{eqnarray}
\frac{f_-(\omega)}{\cosh(c \omega/2)} &=& \chi_+(\omega)+\chi_-(\omega), \nonumber \\
\chi_+(\omega) &=& \frac{2i}{c}\sum_{n=0}^\infty
\frac{(-1)^{n}f_-(-\omega_n)}{\omega+\omega_n}, \nonumber \\
\chi_-(\omega) &=& \frac{f_-(\omega)}{\cosh(c
\omega/2)}-\frac{2i}{c}\sum_{n=0}^\infty
\frac{(-1)^{n}(f_-(-\omega_n))}{\omega+\omega_n}. \label{decomp-A3}
\end{eqnarray}

Applying the formula \eqref{decomp-A3} to \eqref{Fourier} and
\eqref{decomp-A2}, we obtain, for instance,
\begin{equation}
\Phi_+(\omega)=a \frac{i G_-(-i\epsilon)}{\omega+i\epsilon}+
b\frac{i}{c}\sum_{n=0}^\infty
\frac{(-1)^n G_-(-\omega_n)e^{i\lambda_0 \omega_n}}{\omega+\omega_n}.
\end{equation}
Substitution of the above equation and \eqref{pole} into
\eqref{decomp-Phi} then yields
\begin{equation}
\widehat{y}_+(\omega)=G_+(\omega) \left[ a \frac{i
G_-(-i\epsilon)}{\omega+i\epsilon}+ b\frac{i}{c} \frac{ G_-(-\pi
i/c)e^{-\pi \lambda_0/c}}{\omega+\pi i/c}+\cdots \right]. \label{y0}
\end{equation}

\subsection{The case $g(\lambda)=a K(\lambda)$}
Next we consider the case $g(\lambda)=aK(\lambda)$
($a\in\mathbb{C}$). Using the factorization equation
\eqref{factorization}, one finds
\begin{equation}
\widehat{g}(\omega)=a \widehat{K}(\omega)=
a \left(1-\frac{1}{G_+(\omega)G_-(\omega)}\right).
\end{equation}
Thus we immediately obtain $\Phi_\pm(\omega)$ defined by
Eq.~\eqref{decomp-A2}, namely
\begin{equation}
\Phi_+(\omega)=-\frac{a}{G_+(\omega)}, \qquad \Phi_-(\omega)=a
G_-(\omega).
\end{equation}
Then $\widehat{y}_\pm(\omega)$ are respectively given by
\begin{equation}
\widehat{y}_+(\omega)=a (G_+(\omega)-1), \qquad
\widehat{y}_-(\omega)=a \left(1-\frac{1}{G_-(\omega)}\right).
\label{y1}
\end{equation}

\section{Low-temperature behavior for arbitrary $c>0$}\label{low-temp}

In Section~III, we derived the explicit low-temperature
thermodynamics (see \eqref{Free-E} for instance) for the strong
coupling regime ($c\gg 1$) in weak magnetic field ($H\ll 1$) via the
polylogarithm expression for the pressure \eqref{pressure-pl}. Here,
using the dressed function formalism, we extract the universal
low-temperature thermodynamics \eqref{free-E-s} for arbitrary
repulsive coupling $c>0$ in arbitrary magnetic field $H\le H_c$. As
shown in Section~\ref{therm}, the low-temperature thermodynamics is
characterized by the two integral equations
\begin{align}
\varepsilon(k)&=k^2-\mu-\frac{H}{2}-T a_1\ast\ln(1+e^{-\phi_1(k)/T}) \nonumber \\
\phi_1(\lambda)&=H-T a_1\ast \ln(1+e^{-\varepsilon(\lambda)/T})+T
a_2 \ast \ln(1+e^{-\phi_1(\lambda)/T}).
\end{align}
In the limit $T\to 0$, the above equations coincide with the dressed
energies \eqref{dress}.

In completely the same way as in the derivation for
Eqs.~\eqref{phi-0}, \eqref{phi-1} and \eqref{epsilon-1} one obtains
\begin{align}
\varepsilon(k)&=\widetilde{\varepsilon}^{(0)}(k)
+\int_{-\lambda_0}^{\lambda_0}a_1(k-\lambda)\phi_1(\lambda)d \lambda \nonumber \\
\phi_1(\lambda)&=
\widetilde{\phi}_1^{(0)}+\int_{-k_0}^{k_0}a_1(\lambda-k)\varepsilon(k)d
k -\int_{-\lambda_0}^{\lambda_0}a_2(\lambda-\mu)\phi_1(\mu)d \mu
\label{dress-T}
\end{align}
where
\begin{align}
\widetilde{\varepsilon}^{(0)}&=k^2-\mu-\frac{H}{2}- \frac{\pi^2
T^2}{6\varepsilon'_s(\lambda_0)}\left(a_1(k-\lambda_0)+a_1(k+\lambda_0)\right)
\nonumber \\
\widetilde{\phi}_1^{(0)}&=H-\frac{\pi^2 T^2}{6\varepsilon'_c(k_0)}
                     \left(a_1(\lambda-k_0)+a_1(\lambda+k_0)\right)
                   +\frac{\pi^2 T^2}{6\varepsilon'_s(\lambda_0)}
                     \left(a_2(\lambda-\lambda_0)+a_2(\lambda+\lambda_0)\right).
\end{align}
In this limit, the free energy \eqref{Pressure} is reduced to
\begin{equation}
F=\mu n_c-\frac{\pi T^2}{6
\varepsilon_c'(k_0)}+\frac{1}{2\pi}\int_{-k_0}^{k_0}
             \varepsilon(k) dk.
\end{equation}

Applying the dressed function formalism to \eqref{dress-T} and
\eqref{density}, we arrive at
\begin{align}
\frac{1}{2\pi}\int_{-k_0}^{k_0}\varepsilon(k) dk&=
\int_{-k_0}^{k_0} \widetilde{\varepsilon}^{(0)} \rho_c(k) dk+
\int_{-\lambda_0}^{\lambda_0} \widetilde{\phi}_1^{(0)} \rho_s(\lambda) d\lambda \nonumber \\
&=E_0-\frac{\pi T^2}{6}\left(\frac{2\pi
\rho_c(k_0)}{\varepsilon'_c(k_0)}+\frac{2\pi
\rho_s(\lambda_0)}{\varepsilon'_s(\lambda_0)}\right)
     +\frac{\pi T^2}{6\varepsilon_c'(k_0)}-\mu n_c.
\end{align}
This yields
\begin{equation}
F=E_0-\frac{\pi T^2}{6}\left(\frac{1}{v_s}+\frac{1}{v_c}\right)
\label{free-dress}
\end{equation}
where $v_c$ and $v_s$ are, respectively, the holon and spinon
excitation velocities
\begin{equation}
v_c=\frac{\varepsilon_c'(k_0)}{2\pi \rho_c(k_0)}, \qquad
v_s=\frac{\varepsilon_s'(\lambda_0)}{2\pi \rho_s(\lambda_0)}.
\label{fermi-v}
\end{equation}

Before closing this Appendix, we reproduce the low-temperature
thermodynamics \eqref{Free-E} for $c\gg 1$ and $H=0$. At $H=0$, the
Fermi point $\lambda_0=\infty$. By Fourier transformation, the
density functions $\rho_c(k)$ and $\rho_s(\lambda)$ defined by
Eq.~\eqref{density} reduce to
\begin{align}
&\rho_c(k)=\frac{1}{2\pi}+\int_{-k_0}^{k_0}
a_1(k-\lambda)\rho_s(\lambda) d\lambda \nonumber \\
&\rho_s(\lambda)=\int_{-k_0}^{k_0} s(\lambda-k)\rho_c(k) dk.
\label{density-c}
\end{align}
Up to leading order in $1/c$ the second equation reads
\begin{equation}
\rho_s(\lambda)\approx s(\lambda)\int_{-k_0}^{k_0}\rho_c(k) dk=n_c
s(\lambda). \label{rhos}
\end{equation}
Substituting this equation into the first equation in
Eq.~\eqref{density-c} yields
\begin{equation}
\rho_c(k)\approx\frac{1}{2\pi}+n_c
\int_{-\infty}^{\infty}a_1(k-\lambda)s(\lambda) d\lambda
=\frac{1}{2\pi}+n_c K(k)\approx\frac{1}{2\pi}+\frac{\ln 2}{\pi
\gamma} \label{rhoc}
\end{equation}
where \eqref{K} has been used. The dressed energies can be obtained
by just taking the limit $T\to 0$, $H\to 0$ and $\lambda_0\to\infty$
in Eqs~\eqref{denergy} and \eqref{phi-0}. Explicitly,
\begin{align}
&\varepsilon_c(k)\approx k^2-\mu-2\pi P_0 K(k)\approx
k^2-\mu-2P_0\ln 2/c \nonumber
\\ &\varepsilon_s(\lambda)\approx -2\pi P_0 s(\lambda).
\label{energy-cs}
\end{align}

The pressure $P_0$ is determined by solving \eqref{groundstate}.
Combining Eq.~\eqref{rhoc} with Eq.~\eqref{density2} gives the Fermi
point
\begin{equation}
k_0\approx\frac{\pi n_c}{1+\frac{2\ln 2}{\gamma}}\approx \pi
n_c\left(1-\frac{2\ln 2}{\gamma}\right). \label{fermi-p}
\end{equation}
$k_0$ is derived from the condition \eqref{condition}, with result
\begin{equation}
k_0^2=\mu+\frac{2P_0\ln 2}{c}.
\label{chemi}
\end{equation}

Substituting these expressions into \eqref{groundstate} and taking
terms of order $O(1/c)$, one arrives at
\begin{equation}
P_0=-E_0+\mu n_c=\frac{2 n_c^3 \pi^2}{3}\left(1-\frac{6\ln
2}{\gamma}\right) \label{pressure}
\end{equation}
where $\mu$ denotes the zero-temperature chemical potential
determined by Eqs.~\eqref{chemi} and \eqref{fermi-p}:
\begin{equation}
\mu=\pi^2 n_c^2\left(1-\frac{16\ln 2}{3\gamma}\right).
\end{equation}

The excitation velocities
\begin{equation}
v_c\approx 2\pi n_c\left(1-\frac{4\ln2}\gamma \right), \qquad
v_s\approx\frac{2\pi^3 n_c}{3\gamma}\left(1-\frac{6\ln
2}{\gamma}\right). \label{fermi-h0}
\end{equation}
are evaluated by substituting all the results into
Eq.~\eqref{fermi-v}. Thus from Eq.~\eqref{free-dress}, one finds
that the low-temperature free energy $F$ agrees with
Eq.~\eqref{Free-E} at $H=0$.

\section{Correlation functions for   $H\ll 1$} \label{CF-1}

Here we consider the leading terms of the zero temperature correlation
functions listed in Section V. The conformal dimensions are
\begin{eqnarray}
\nonumber \lefteqn{2\Delta_{c}^{\pm}(\Delta
N_{c,s},N^{\pm}_{c,s},D_{c,s})=}
\\ && \nonumber \left(D_{c}+\frac{1}{2}D_{s}\pm\frac{1}{2}\Delta
N_{c}\right)^{2}-\frac{4}{\pi^{2}}\left(\frac{H}{H_{c}}\right)D_{s}\left(D_{c}+\frac{1}{2}D_{s}\pm\frac{1}{2}\Delta
N_{c}\right) \\ && \nonumber +\frac{4\ln
2}{\gamma}\left[\left(D_{c}+\frac{1}{2}D_{s}\right)^{2}-\frac{1}{4}(\Delta
N_{c})^{2}\right]-\frac{8\ln
2}{\pi^{2}\gamma}\left(\frac{H}{H_{c}}\right)D_{s}\left(D_{c}+\frac{1}{2}D_{s}\mp\frac{1}{2}\Delta
N_{c}\right) \\ &&
\pm\frac{4}{\gamma}\left(\frac{H}{H_{c}}\right)\left(\frac{\Delta
N_{c}}{2}-\Delta
N_{s}\right)\left(D_{c}+\frac{1}{2}D_{s}\pm\frac{1}{2}\Delta
N_{c}\right)+2N^{\pm}_{c}
\end{eqnarray}
and
\begin{eqnarray}
\nonumber \lefteqn{2\Delta_{s}^{\pm}(\Delta
N_{c,s},N^{\pm}_{c,s},D_{c,s})=} \\ && \nonumber
\frac{1}{2}\left(\Delta N_{s}-\frac{1}{2}\Delta N_{c}\pm
D_{s}\right)^{2}
+\frac{2}{\pi^{2}}\left(\frac{H}{H_{c}}\right)\Delta
N_{c}\left(\Delta N_{s}-\frac{1}{2}\Delta N_{c}\pm D_{s}\right)\\ &&
\nonumber +\frac{1}{4\ln(H_{0}/H)}\left[D_{s}^{2}-\left(\Delta
N_{s}-\frac{1}{2}\Delta
N_{c}\right)^{2}\right]\pm\frac{4}{\gamma}\left(\frac{H}{H_{c}}\right)\left(D_{c}+\frac{1}{2}D_{s}\right)\left(\Delta
N_{s}-\frac{1}{2}\Delta N_{c}\pm D_{s}\right) \\ && -\frac{4\ln
2}{\pi^{2}\gamma}\left(\frac{H}{H_{c}}\right)\Delta
N_{c}\left(\Delta N_{s}-\frac{1}{2}\Delta N_{c}\pm D_{s}\right)
+2N^{\pm}_{s}.
\end{eqnarray}

(i) The leading orders for the field correlator $G_{\uparrow}(x,t)$
come from the set of quantum numbers $(D_{c},D_{s})=(1/2,-1/2)$ and
$(1/2,1/2)$. The conformal dimensions corresponding to these sets of
quantum numbers are
\begin{eqnarray}
&& \nonumber
2\Delta_{c}^{+}=\frac{9}{16}-\frac{3\ln
2}{4\gamma}+\frac{3}{2\pi^{2}}\left(\frac{H}{H_{c}}\right)+\frac{3}{2\gamma}\left(\frac{H}{H_{c}}\right)
-\frac{\ln 2}{\pi^{2}\gamma}\left(\frac{H}{H_{c}}\right) \\
&& \nonumber 2\Delta_{c}^{-}=\frac{1}{16}-\frac{3\ln
2}{4\gamma}-\frac{1}{2\pi^{2}}\left(\frac{H}{H_{c}}\right)+\frac{1}{2\gamma}\left(\frac{H}{H_{c}}\right)
+\frac{3\ln 2}{\pi^{2}\gamma}\left(\frac{H}{H_{c}}\right)
\\ && \nonumber
2\Delta_{s}^{+}=\frac{1}{2}-\frac{2}{\pi^{2}}\left(\frac{H}{H_{c}}\right)-\frac{1}{\gamma}\left(\frac{H}{H_{c}}\right)
+\frac{4\ln 2}{\pi^{2}\gamma}\left(\frac{H}{H_{c}}\right)
\\ && 2\Delta_{s}^{-}=0
\end{eqnarray}
for $(D_{c},D_{s})=(1/2,-1/2)$ and
\begin{eqnarray}
&& \nonumber
2\Delta_{c}^{+}=\frac{25}{16}+\frac{5\ln
2}{4\gamma}-\frac{5}{2\pi^{2}}\left(\frac{H}{H_{c}}\right)+\frac{5}{2\gamma}\left(\frac{H}{H_{c}}\right)
-\frac{\ln 2}{\pi^{2}\gamma}\left(\frac{H}{H_{c}}\right)
 \\ && \nonumber 2\Delta_{c}^{-}=\frac{1}{16}+\frac{5\ln
2}{4\gamma}-\frac{1}{2\pi^{2}}\left(\frac{H}{H_{c}}\right)-\frac{1}{2\gamma}\left(\frac{H}{H_{c}}\right)
-\frac{5\ln 2}{\pi^{2}\gamma}\left(\frac{H}{H_{c}}\right)
\\ && \nonumber 2\Delta_{s}^{+}=0 \\ &&
2\Delta_{s}^{-}=\frac{1}{2}-\frac{2}{\pi^{2}}\left(\frac{H}{H_{c}}\right)+\frac{3}{\gamma}\left(\frac{H}{H_{c}}\right)
+\frac{4\ln 2}{\pi^{2}\gamma}\left(\frac{H}{H_{c}}\right)
\end{eqnarray}
for $(D_{c},D_{s})=(1/2,1/2)$.
Hence we obtain
\begin{equation}
G_{\uparrow}(x,t)\approx
\frac{A_{1}\cos(k_{F\uparrow}x)}{|x+iv_{c}t|^{\theta_{c1}}|x+iv_{s}t|^{\theta_{s1}}}
+\frac{A_{2}\cos((k_{F\uparrow}+2k_{F\downarrow})x)}{|x+iv_{c}t|^{\theta_{c2}}|x+iv_{s}t|^{\theta_{s2}}}
\end{equation}
where the exponents are given by
\begin{eqnarray}
\nonumber \theta_{c1} &=& \frac{5}{8}-\frac{3\ln
2}{2\gamma}+\frac{1}{\pi^{2}}\left(\frac{H}{H_{c}}\right)+\frac{2}{\gamma}\left(\frac{H}{H_{c}}\right)
+\frac{2\ln 2}{\pi^{2}\gamma}\left(\frac{H}{H_{c}}\right) \\
\nonumber \theta_{c2} &=& \frac{13}{8}+\frac{5\ln
2}{2\gamma}-\frac{3}{\pi^{2}}\left(\frac{H}{H_{c}}\right)+\frac{2}{\gamma}\left(\frac{H}{H_{c}}\right)
-\frac{6\ln 2}{\pi^{2}\gamma}\left(\frac{H}{H_{c}}\right) \\
\nonumber \theta_{s1} &=&
\frac{1}{2}-\frac{2}{\pi^{2}}\left(\frac{H}{H_{c}}\right)-\frac{1}{\gamma}\left(\frac{H}{H_{c}}\right)
+\frac{4\ln 2}{\pi^{2}\gamma}\left(\frac{H}{H_{c}}\right) \\
\theta_{s2} &=&
\frac{1}{2}-\frac{2}{\pi^{2}}\left(\frac{H}{H_{c}}\right)+\frac{3}{\gamma}\left(\frac{H}{H_{c}}\right)
+\frac{4\ln 2}{\pi^{2}\gamma}\left(\frac{H}{H_{c}}\right).
\end{eqnarray}

(ii) The leading terms for the field correlator
$G_{\downarrow}(x,t)$ come from the set of quantum numbers
$(D_{c},D_{s})=(0,1/2)$ and $(1,-1/2)$. The corresponding conformal
dimensions are
\begin{eqnarray}
&& \nonumber
2\Delta_{c}^{+}=\frac{9}{16}-\frac{3\ln
2}{4\gamma}-\frac{3}{2\pi^{2}}\left(\frac{H}{H_{c}}\right)-\frac{3}{2\gamma}\left(\frac{H}{H_{c}}\right)
+\frac{\ln 2}{\pi^{2}\gamma}\left(\frac{H}{H_{c}}\right)
 \\
&& \nonumber 2\Delta_{c}^{-}=\frac{1}{16}-\frac{3\ln
2}{4\gamma}+\frac{1}{2\pi^{2}}\left(\frac{H}{H_{c}}\right)-\frac{1}{2\gamma}\left(\frac{H}{H_{c}}\right)
-\frac{3\ln 2}{\pi^{2}\gamma}\left(\frac{H}{H_{c}}\right)
\\ && \nonumber
2\Delta_{s}^{+}=\frac{1}{2}+\frac{2}{\pi^{2}}\left(\frac{H}{H_{c}}\right)+\frac{1}{\gamma}\left(\frac{H}{H_{c}}\right)
-\frac{4\ln 2}{\pi^{2}\gamma}\left(\frac{H}{H_{c}}\right)
\\ && 2\Delta_{s}^{-}=0
\end{eqnarray}
for $(D_{c},D_{s})=(0,1/2)$ and
\begin{eqnarray}
&& \nonumber
2\Delta_{c}^{+}=\frac{25}{16}+\frac{5\ln
2}{4\gamma}+\frac{5}{2\pi^{2}}\left(\frac{H}{H_{c}}\right)-\frac{5}{2\gamma}\left(\frac{H}{H_{c}}\right)
+\frac{\ln 2}{\pi^{2}\gamma}\left(\frac{H}{H_{c}}\right)
\\ && \nonumber 2\Delta_{c}^{-}=\frac{1}{16}+\frac{5\ln
2}{4\gamma}+\frac{1}{2\pi^{2}}\left(\frac{H}{H_{c}}\right)+\frac{1}{2\gamma}\left(\frac{H}{H_{c}}\right)
+\frac{5\ln 2}{\pi^{2}\gamma}\left(\frac{H}{H_{c}}\right)
\\ && \nonumber 2\Delta_{s}^{+}=0 \\ &&
2\Delta_{s}^{-}=\frac{1}{2}+\frac{2}{\pi^{2}}\left(\frac{H}{H_{c}}\right)-\frac{3}{\gamma}\left(\frac{H}{H_{c}}\right)
-\frac{4\ln 2}{\pi^{2}\gamma}\left(\frac{H}{H_{c}}\right)
\end{eqnarray}
for $(D_{c},D_{s})=(1,-1/2)$.
This gives
\begin{equation}
G_{\downarrow}(x,t)\approx
\frac{A_{1}\cos(k_{F\downarrow}x)}{|x+iv_{c}t|^{\theta_{c1}}|x+iv_{s}t|^{\theta_{s1}}}
+\frac{A_{2}\cos((2k_{F\uparrow}+k_{F\downarrow})x)}{|x+iv_{c}t|^{\theta_{c2}}|x+iv_{s}t|^{\theta_{s2}}}
\end{equation}
where
\begin{eqnarray}
\nonumber \theta_{c1} &=& \frac{5}{8}-\frac{3\ln
2}{2\gamma}-\frac{1}{\pi^{2}}\left(\frac{H}{H_{c}}\right)-\frac{2}{\gamma}\left(\frac{H}{H_{c}}\right)
-\frac{2\ln 2}{\pi^{2}\gamma}\left(\frac{H}{H_{c}}\right) \\
\nonumber \theta_{c2} &=& \frac{13}{8}+\frac{5\ln
2}{2\gamma}+\frac{3}{\pi^{2}}\left(\frac{H}{H_{c}}\right)-\frac{2}{\gamma}\left(\frac{H}{H_{c}}\right)
+\frac{6\ln 2}{\pi^{2}\gamma}\left(\frac{H}{H_{c}}\right) \\
\nonumber \theta_{s1} &=&
\frac{1}{2}+\frac{2}{\pi^{2}}\left(\frac{H}{H_{c}}\right)+\frac{1}{\gamma}\left(\frac{H}{H_{c}}\right)
-\frac{4\ln 2}{\pi^{2}\gamma}\left(\frac{H}{H_{c}}\right) \\
\theta_{s2} &=&
\frac{1}{2}+\frac{2}{\pi^{2}}\left(\frac{H}{H_{c}}\right)-\frac{3}{\gamma}\left(\frac{H}{H_{c}}\right)
-\frac{4\ln 2}{\pi^{2}\gamma}\left(\frac{H}{H_{c}}\right).
\end{eqnarray}

(iii) The leading terms for the charge density correlator come from
the set of quantum numbers $(D_{c},D_{s})=(0,0)$, $(0,1)$, $(1,0)$
and $(1,-1)$. The corresponding conformal dimensions are
\begin{eqnarray}
&& \nonumber 2\Delta_{c}^{+}=0, \quad
2\Delta_{s}^{+}=0, \\ && 2\Delta_{c}^{-}=0, \quad 2\Delta_{s}^{-}=0,
\end{eqnarray}
for $(D_{c},D_{s})=(0,0)$ and
\begin{eqnarray}
&& \nonumber
2\Delta_{c}^{+}=\frac{1}{4}+\frac{\ln
2}{\gamma}-\frac{2}{\pi^{2}}\left(\frac{H}{H_{c}}\right)-\frac{4\ln
2}{\pi^{2}\gamma}\left(\frac{H}{H_{c}}\right)
\\ && \nonumber
2\Delta_{c}^{-}=\frac{1}{4}+\frac{\ln
2}{\gamma}-\frac{2}{\pi^{2}}\left(\frac{H}{H_{c}}\right)-\frac{4\ln
2}{\pi^{2}\gamma}\left(\frac{H}{H_{c}}\right) \\ && \nonumber
2\Delta_{s}^{+}=\frac{1}{2}+\frac{1}{4\ln(H_{0}/H)}+\frac{2}{\gamma}\left(\frac{H}{H_{c}}\right)
\\ && 2\Delta_{s}^{-}=\frac{1}{2}+\frac{1}{4\ln(H_{0}/H)}+\frac{2}{\gamma}\left(\frac{H}{H_{c}}\right).
\end{eqnarray}
for $(D_{c},D_{s})=(0,1)$ with
\begin{eqnarray}
&& \nonumber 2\Delta_{c}^{+}=1+\frac{4\ln
2}{\gamma}, \quad 2\Delta_{s}^{+}=0, \\ &&
2\Delta_{c}^{-}=1+\frac{4\ln 2}{\gamma}, \quad 2\Delta_{s}^{-}=0.
\end{eqnarray}
for $(D_{c},D_{s})=(1,0)$ and
\begin{eqnarray}
&& \nonumber
2\Delta_{c}^{+}=\frac{1}{4}+\frac{\ln
2}{\gamma}+\frac{2}{\pi^{2}}\left(\frac{H}{H_{c}}\right)+\frac{4\ln
2}{\pi^{2}\gamma}\left(\frac{H}{H_{c}}\right)
 \\ && \nonumber
2\Delta_{c}^{-}=\frac{1}{4}+\frac{\ln
2}{\gamma}+\frac{2}{\pi^{2}}\left(\frac{H}{H_{c}}\right)+\frac{4\ln
2}{\pi^{2}\gamma}\left(\frac{H}{H_{c}}\right) \\ && \nonumber
2\Delta_{s}^{+}=\frac{1}{2}+\frac{1}{4\ln(H_{0}/H)}-\frac{2}{\gamma}\left(\frac{H}{H_{c}}\right)
\\ && 2\Delta_{s}^{-}=\frac{1}{2}+\frac{1}{4\ln(H_{0}/H)}-\frac{2}{\gamma}\left(\frac{H}{H_{c}}\right).
\end{eqnarray}
for $(D_{c},D_{s})=(1,-1)$.
The correlation function is then given by
\begin{equation}
G_{nn}(x,t)\approx n^{2}
+\frac{A_{1}\cos(2k_{F\downarrow}x)}{|x+iv_{c}t|^{\theta_{c1}}|x+iv_{s}t|^{\theta_{s1}}}
+\frac{A_{2}\cos(2k_{F\uparrow}x)}{|x+iv_{c}t|^{\theta_{c2}}|x+iv_{s}t|^{\theta_{s2}}}
+\frac{A_{3}\cos(2(k_{F\downarrow}+k_{F\uparrow})x)}{|x+iv_{c}t|^{\theta_{c3}}}
\end{equation}
where
\begin{eqnarray}
\nonumber \theta_{c1} &=& \frac{1}{2}+\frac{2\ln
2}{\gamma}-\frac{4}{\pi^{2}}\left(\frac{H}{H_{c}}\right)-\frac{8\ln
2}{\pi^{2}\gamma}\left(\frac{H}{H_{c}}\right) \\
\nonumber \theta_{c2} &=& \frac{1}{2}+\frac{2\ln
2}{\gamma}+\frac{4}{\pi^{2}}\left(\frac{H}{H_{c}}\right)+\frac{8\ln
2}{\pi^{2}\gamma}\left(\frac{H}{H_{c}}\right) \\ \nonumber
\theta_{c3} &=& 2+\frac{8\ln 2}{\gamma} \\ \nonumber \theta_{s1} &=&
1+\frac{1}{2\ln(H_{0}/H)}+\frac{4}{\gamma}\left(\frac{H}{H_{c}}\right)
\\ \theta_{s2} &=& 1+\frac{1}{2\ln(H_{0}/H)}-\frac{4}{\gamma}\left(\frac{H}{H_{c}}\right).
\end{eqnarray}

(iv) The longitudinal spin-spin correlator is similar to the charge
density correlator obtained above. The only difference is
that the leading term $n^{2}$ is replaced by $(m^{z})^{2}$.

(v) The leading terms of the transverse spin-spin correlator are
obtained from the quantum numbers $(D_{c},D_{s})=(1/2,0)$ and
$(1/2,-1)$. The corresponding conformal dimensions are
\begin{eqnarray}
&& \nonumber
2\Delta_{c}^{+}=\frac{1}{4}+\frac{\ln
2}{\gamma}-\frac{2}{\gamma}\left(\frac{H}{H_{c}}\right)
 \\ && \nonumber
2\Delta_{c}^{-}=\frac{1}{4}+\frac{\ln
2}{\gamma}+\frac{2}{\gamma}\left(\frac{H}{H_{c}}\right) \\ &&
\nonumber 2\Delta_{s}^{+}=\frac{1}{2}-\frac{1}{4\ln(H_{0}/H)}+\frac{2}{\gamma}\left(\frac{H}{H_{c}}\right) \\
&&
2\Delta_{s}^{-}=\frac{1}{2}-\frac{1}{4\ln(H_{0}/H)}-\frac{2}{\gamma}\left(\frac{H}{H_{c}}\right)
\end{eqnarray}
for $(D_{c},D_{s})=(1/2,0)$ and
\begin{eqnarray}
&& \nonumber 2\Delta_{c}^{+}=0, \quad
2\Delta_{s}^{+}=0, \\ && 2\Delta_{c}^{-}=0, \quad 2\Delta_{s}^{-}=2
\end{eqnarray}
for $(D_{c},D_{s})=(1/2,-1)$.
The correlation function is
\begin{equation}
G^{\perp}(x,t)\approx
\frac{A_{1}\cos((k_{F\downarrow}+k_{F\uparrow})x)}{|x+iv_{c}t|^{\theta_{c1}}|x+iv_{s}t|^{\theta_{s1}}}
+\frac{A_{2}\cos((k_{F\uparrow}-k_{F\downarrow})x)}{|x+iv_{s}t|^{\theta_{s2}}}
\end{equation}
where
\begin{eqnarray}
\nonumber \theta_{c1} &=& \frac{1}{2}+\frac{2\ln
2}{\gamma} \\
\nonumber \theta_{s1} &=&
1-\frac{1}{2\ln(H_{0}/H)}+\frac{4}{\gamma}\left(\frac{H}{H_{c}}\right)
\\ \theta_{s2} &=& 2.
\end{eqnarray}

(vi) Lastly we consider the pair correlator. The leading order
terms are contributed from the quantum numbers
$(D_{c},D_{s})=(1/2,0)$ and $(1/2,-1)$. The corresponding conformal
dimensions are
\begin{eqnarray}
&& \nonumber
2\Delta_{c}^{+}=\frac{9}{4}-\frac{3\ln 2}{\gamma}, \quad 2\Delta_{s}^{+}=0, \\
&& 2\Delta_{c}^{-}=\frac{1}{4}-\frac{3\ln 2}{\gamma}, \quad
2\Delta_{s}^{-}=0,
\end{eqnarray}
for $(D_{c},D_{s})=(1/2,0)$ and
\begin{eqnarray}
&& \nonumber
2\Delta_{c}^{+}=1-\frac{4\ln
2}{\gamma}+\frac{4}{\pi^{2}}\left(\frac{H}{H_{c}}\right)-\frac{8\ln
2}{\pi^{2}\gamma}\left(\frac{H}{H_{c}}\right)
\\ && \nonumber 2\Delta_{c}^{-}=1-\frac{4\ln
2}{\gamma}-\frac{4}{\pi^{2}}\left(\frac{H}{H_{c}}\right)+\frac{8\ln
2}{\pi^{2}\gamma}\left(\frac{H}{H_{c}}\right)
\\ && \nonumber
2\Delta_{s}^{+}=\frac{1}{2}-\frac{4}{\pi^{2}}\left(\frac{H}{H_{c}}\right)+\frac{1}{4\ln(H_{0}/H)}
+\frac{8\ln 2}{\pi^{2}\gamma}\left(\frac{H}{H_{c}}\right)
\\ &&
2\Delta_{s}^{-}=\frac{1}{2}+\frac{4}{\pi^{2}}\left(\frac{H}{H_{c}}\right)+\frac{1}{4\ln(H_{0}/H)}
-\frac{8\ln 2}{\pi^{2}\gamma}\left(\frac{H}{H_{c}}\right)
\end{eqnarray}
for $(D_{c},D_{s})=(1/2,-1)$.
The correlation function is then given by
\begin{equation}
G_{p}(x,t)\approx
\frac{A_{1}\cos((k_{F\downarrow}+k_{F\uparrow})x)}{|x+iv_{c}t|^{\theta_{c1}}}
+\frac{A_{2}\cos((k_{F\uparrow}-k_{F\downarrow})x)}{|x+iv_{c}t|^{\theta_{c2}}|x+iv_{s}t|^{\theta_{s1}}}
\end{equation}
where
\begin{eqnarray}
\nonumber \theta_{c1} &=& \frac{5}{2}-\frac{6\ln
2}{\gamma} \\
\nonumber \theta_{c2} &=& 2-\frac{8\ln
2}{\gamma} \\
\theta_{s1} &=& 1+\frac{1}{2\ln(H_{0}/H)}.
\end{eqnarray}

We have obtained the charge and spin velocities in Eq.~(\ref{fermi-h0})
for this regime. An extension to the finite temperature correlation
functions is straightforward.

\section{Correlation functions for   $H\rightarrow H_{c}$} \label{CF-2}

The conformal dimensions in this case are
\begin{eqnarray}
\nonumber \lefteqn{2\Delta_{c}^{\pm}(\Delta
N_{c,s},N^{\pm}_{c,s},D_{c,s})=} \\
&& \nonumber \left(D_{c}\pm\frac{1}{2}\Delta
N_{c}\right)^{2}+2\left(D_{c}\pm\frac{1}{2}\Delta
N_{c}\right)\left(\frac{2}{\pi}\sqrt{1-\frac{H}{H_{c}}}D_{s}\mp\frac{2}{\gamma}\Delta
N_{s}+\frac{8}{\pi\gamma}\sqrt{1-\frac{H}{H_{c}}}D_{c}\right)
\\ && \mp\frac{8}{\pi\gamma}\sqrt{1-\frac{H}{H_{c}}}D_{s}\Delta
N_{s}+2N^{\pm}_{c}
\end{eqnarray}
and
\begin{eqnarray}
\nonumber \lefteqn{2\Delta_{s}^{\pm}(\Delta
N_{c,s},N^{\pm}_{c,s},D_{c,s})=} \\
&& \nonumber \left(D_{s}\pm\frac{1}{2}\Delta
N_{s}\right)^{2}+2\left(D_{s}\pm\frac{1}{2}\Delta
N_{s}\right)\left[\frac{4}{\gamma}\left(1-\frac{1}{\pi^{2}}\sqrt{1-\frac{H}{H_{c}}}\right)D_{c}\right.
\\ && \nonumber
\left.-\frac{1}{\pi}\sqrt{1-\frac{H}{H_{c}}}\left(D_{s}-\frac{8}{\gamma}D_{s}\pm\left[\Delta
N_{c}-\frac{1}{2}\Delta N_{s}\right]\right)\right]
\\ && -\frac{8}{\pi\gamma}\sqrt{1-\frac{H}{H_{c}}}D_{c}\left(D_{s}\pm\left[\Delta
N_{c}-\frac{1}{2}\Delta N_{s}\right]\right)+2N^{\pm}_{s}.
\end{eqnarray}

(i) $G_{\uparrow}(x,t)$: The conformal dimensions are
\begin{eqnarray}
&& \nonumber
2\Delta_{c}^{+}=1-\frac{2}{\pi}\sqrt{1-\frac{H}{H_{c}}}+\frac{8}{\pi\gamma}\sqrt{1-\frac{H}{H_{c}}}
\\ && \nonumber 2\Delta_{c}^{-}=0 \\ && \nonumber
2\Delta_{s}^{+}=\frac{1}{4}-\frac{2}{\gamma}+\frac{1}{2\pi}\sqrt{1-\frac{H}{H_{c}}}+\frac{4}{\pi\gamma}\sqrt{1-\frac{H}{H_{c}}}
\\ &&
2\Delta_{s}^{-}=\frac{1}{4}-\frac{2}{\gamma}-\frac{3}{2\pi}\sqrt{1-\frac{H}{H_{c}}}+\frac{12}{\pi\gamma}\sqrt{1-\frac{H}{H_{c}}}
\end{eqnarray}
for $(D_{c},D_{s})=(1/2,-1/2)$ and
\begin{eqnarray}
&& \nonumber
2\Delta_{c}^{+}=1+\frac{2}{\pi}\sqrt{1-\frac{H}{H_{c}}}+\frac{8}{\pi\gamma}\sqrt{1-\frac{H}{H_{c}}}
\\ && \nonumber 2\Delta_{c}^{-}=0 \\ && \nonumber
2\Delta_{s}^{+}=\frac{1}{4}+\frac{2}{\gamma}-\frac{3}{2\pi}\sqrt{1-\frac{H}{H_{c}}}
-\frac{4}{\pi\gamma}\sqrt{1-\frac{H}{H_{c}}}
\\ &&
2\Delta_{s}^{-}=\frac{1}{4}+\frac{2}{\gamma}+\frac{1}{2\pi}\sqrt{1-\frac{H}{H_{c}}}
+\frac{4}{\pi\gamma}\sqrt{1-\frac{H}{H_{c}}}
\end{eqnarray}
for $(D_{c},D_{s})=(1/2,1/2)$.
The correlation function is
\begin{equation}
G_{\uparrow}(x,t)\approx
\frac{A_{1}\cos(2k_{F\uparrow}x)}{|x+iv_{c}t|^{\theta_{c1}}|x+iv_{s}t|^{\theta_{s1}}}
+\frac{A_{2}\cos((k_{F\uparrow}+2k_{F\downarrow})x)}{|x+iv_{c}t|^{\theta_{c2}}|x+iv_{s}t|^{\theta_{s2}}}
\end{equation}
with
\begin{eqnarray}
\nonumber \theta_{c1} &=&
1-\frac{2}{\pi}\sqrt{1-\frac{H}{H_{c}}}+\frac{8}{\pi\gamma}\sqrt{1-\frac{H}{H_{c}}}
\\ \nonumber
\theta_{c2} &=&
1+\frac{2}{\pi}\sqrt{1-\frac{H}{H_{c}}}+\frac{8}{\pi\gamma}\sqrt{1-\frac{H}{H_{c}}}
\\ \nonumber
\theta_{s1} &=&
\frac{1}{2}-\frac{4}{\gamma}-\frac{1}{\pi}\sqrt{1-\frac{H}{H_{c}}}+\frac{16}{\pi\gamma}\sqrt{1-\frac{H}{H_{c}}}
\\ \theta_{s2} &=&
\frac{1}{2}+\frac{4}{\gamma}-\frac{1}{\pi}\sqrt{1-\frac{H}{H_{c}}}.
\end{eqnarray}

(ii) $G_{\downarrow}(x,t)$: The conformal dimensions are
\begin{eqnarray}
&& \nonumber
2\Delta_{c}^{+}=\frac{1}{4}-\frac{2}{\gamma}+\frac{1}{\pi}\sqrt{1-\frac{H}{H_{c}}}
-\frac{4}{\pi\gamma}\sqrt{1-\frac{H}{H_{c}}}
\\ && \nonumber
2\Delta_{c}^{-}=\frac{1}{4}-\frac{2}{\gamma}-\frac{1}{\pi}\sqrt{1-\frac{H}{H_{c}}}
+\frac{4}{\pi\gamma}\sqrt{1-\frac{H}{H_{c}}}
\\ && \nonumber 2\Delta_{s}^{+}=1-\frac{2}{\pi}\sqrt{1-\frac{H}{H_{c}}}+\frac{8}{\pi\gamma}\sqrt{1-\frac{H}{H_{c}}} \\ &&
2\Delta_{s}^{-}=0
\end{eqnarray}
for $(D_{c},D_{s})=(0,1/2)$ and
\begin{eqnarray}
&& \nonumber
2\Delta_{c}^{+}=\frac{9}{4}-\frac{6}{\gamma}-\frac{3}{\pi}\sqrt{1-\frac{H}{H_{c}}}
+\frac{28}{\pi\gamma}\sqrt{1-\frac{H}{H_{c}}}
\\ && \nonumber
2\Delta_{c}^{-}=\frac{1}{4}+\frac{2}{\gamma}-\frac{1}{\pi}\sqrt{1-\frac{H}{H_{c}}}
+\frac{4}{\pi\gamma}\sqrt{1-\frac{H}{H_{c}}}
\\ && \nonumber 2\Delta_{s}^{+}=0 \\ &&
2\Delta_{s}^{-}=1-\frac{8}{\gamma}-\frac{2}{\pi}\sqrt{1-\frac{H}{H_{c}}}
+\frac{24}{\pi\gamma}\sqrt{1-\frac{H}{H_{c}}}
\end{eqnarray}
for $(D_{c},D_{s})=(1,-1/2)$.
The correlation function is
\begin{equation}
G_{\downarrow}(x,t)\approx
\frac{A_{1}\cos(2k_{F\downarrow}x)}{|x+iv_{c}t|^{\theta_{c1}}|x+iv_{s}t|^{\theta_{s1}}}
+\frac{A_{2}\cos((2k_{F\uparrow}+k_{F\downarrow})x)}{|x+iv_{c}t|^{\theta_{c2}}|x+iv_{s}t|^{\theta_{s2}}}
\end{equation}
where
\begin{eqnarray}
\nonumber \theta_{c1} &=& \frac{1}{2}-\frac{4}{\gamma} \\ \nonumber
\theta_{c2} &=&
\frac{5}{2}-\frac{4}{\gamma}-\frac{4}{\pi}\sqrt{1-\frac{H}{H_{c}}}+\frac{32}{\pi\gamma}\sqrt{1-\frac{H}{H_{c}}}
\\ \nonumber
\theta_{s1} &=& 1-\frac{2}{\pi}\sqrt{1-\frac{H}{H_{c}}}+\frac{8}{\pi\gamma}\sqrt{1-\frac{H}{H_{c}}} \\
\theta_{s2} &=&
1-\frac{8}{\gamma}-\frac{2}{\pi}\sqrt{1-\frac{H}{H_{c}}}+\frac{24}{\pi\gamma}\sqrt{1-\frac{H}{H_{c}}}.
\end{eqnarray}

(iii) $G_{nn}$: The conformal dimensions are
\begin{eqnarray}
&& \nonumber 2\Delta_{c}^{+}=0, \quad
2\Delta_{s}^{+}=0, \\ && 2\Delta_{c}^{-}=0, \quad 2\Delta_{s}^{-}=0
\end{eqnarray}
for $(D_{c},D_{s})=(0,0)$ and
\begin{eqnarray}
&& \nonumber 2\Delta_{c}^{+}=0, \quad
2\Delta_{s}^{+}=1-\frac{2}{\pi}\sqrt{1-\frac{H}{H_{c}}}+\frac{16}{\pi\gamma}\sqrt{1-\frac{H}{H_{c}}},
\\ && 2\Delta_{c}^{-}=0, \quad
2\Delta_{s}^{-}=1-\frac{2}{\pi}\sqrt{1-\frac{H}{H_{c}}}+\frac{16}{\pi\gamma}\sqrt{1-\frac{H}{H_{c}}}
\end{eqnarray}
for $(D_{c},D_{s})=(0,1)$ with
\begin{eqnarray}
&& \nonumber
2\Delta_{c}^{+}=1+\frac{16}{\pi\gamma}\sqrt{1-\frac{H}{H_{c}}},
\quad 2\Delta_{s}^{+}=0, \\ &&
2\Delta_{c}^{-}=1+\frac{16}{\pi\gamma}\sqrt{1-\frac{H}{H_{c}}},
\quad 2\Delta_{s}^{-}=0
\end{eqnarray}
for $(D_{c},D_{s})=(1,0)$ and
\begin{eqnarray}
&& \nonumber
2\Delta_{c}^{+}=1-\frac{4}{\pi}\sqrt{1-\frac{H}{H_{c}}}+\frac{16}{\pi\gamma}\sqrt{1-\frac{H}{H_{c}}}
\\ && \nonumber
2\Delta_{c}^{-}=1-\frac{4}{\pi}\sqrt{1-\frac{H}{H_{c}}}+\frac{16}{\pi\gamma}\sqrt{1-\frac{H}{H_{c}}}
\\ && \nonumber
2\Delta_{s}^{+}=1-\frac{8}{\gamma}-\frac{2}{\pi}\sqrt{1-\frac{H}{H_{c}}}+\frac{32}{\pi\gamma}\sqrt{1-\frac{H}{H_{c}}}
\\ &&
2\Delta_{s}^{-}=1-\frac{8}{\gamma}-\frac{2}{\pi}\sqrt{1-\frac{H}{H_{c}}}+\frac{32}{\pi\gamma}\sqrt{1-\frac{H}{H_{c}}}
\end{eqnarray}
for $(D_{c},D_{s})=(1,-1)$.
The correlation function is then given by
\begin{equation}
G_{nn}(x,t)\approx n^{2}
+\frac{A_{1}\cos(2k_{F\downarrow}x)}{|x+iv_{s}t|^{\theta_{s1}}}
+\frac{A_{2}\cos(2k_{F\uparrow}x)}{|x+iv_{c}t|^{\theta_{c1}}|x+iv_{s}t|^{\theta_{s2}}}
+\frac{A_{3}\cos(2(k_{F\downarrow}+k_{F\uparrow})x)}{|x+iv_{c}t|^{\theta_{c2}}}
\end{equation}
where
\begin{eqnarray}
\nonumber \theta_{c1} &=&
2-\frac{8}{\pi}\sqrt{1-\frac{H}{H_{c}}}+\frac{32}{\pi\gamma}\sqrt{1-\frac{H}{H_{c}}}
\\ \nonumber
\theta_{c2} &=& 2+\frac{32}{\pi\gamma}\sqrt{1-\frac{H}{H_{c}}} \\
\nonumber \theta_{s1} &=&
2-\frac{4}{\pi}\sqrt{1-\frac{H}{H_{c}}}+\frac{32}{\pi\gamma}\sqrt{1-\frac{H}{H_{c}}} \\
\theta_{s2} &=&
2-\frac{16}{\gamma}-\frac{4}{\pi}\sqrt{1-\frac{H}{H_{c}}}+\frac{64}{\pi\gamma}\sqrt{1-\frac{H}{H_{c}}}.
\end{eqnarray}

(iv) $G^{z}(x,t)$ has the same form as $G_{nn}(x,t)$ except
the term $n^{2}$ is replaced by $(m^{z})^{2}$.

(v) $G^{\perp}(x,t)$: The conformal dimensions are
\begin{eqnarray}
&& \nonumber
2\Delta_{c}^{+}=\frac{1}{4}-\frac{2}{\gamma}+\frac{4}{\pi\gamma}\sqrt{1-\frac{H}{H_{c}}}
\\ && \nonumber 2\Delta_{c}^{-}=\frac{1}{4}+\frac{2}{\gamma}+\frac{4}{\pi\gamma}\sqrt{1-\frac{H}{H_{c}}} \\ &&
\nonumber
2\Delta_{s}^{+}=\frac{1}{4}+\frac{2}{\gamma}+\frac{1}{2\pi}\sqrt{1-\frac{1}{H_{c}}}
\\ &&
2\Delta_{s}^{-}=\frac{1}{4}-\frac{2}{\gamma}+\frac{1}{2\pi}\sqrt{1-\frac{1}{H_{c}}}
\end{eqnarray}
for $(D_{c},D_{s})=(1/2,0)$ and
\begin{eqnarray}
&& \nonumber
2\Delta_{c}^{+}=\frac{1}{4}-\frac{2}{\gamma}-\frac{2}{\pi}\sqrt{1-\frac{1}{H_{c}}}+\frac{12}{\pi\gamma}\sqrt{1-\frac{H}{H_{c}}}
\\ && \nonumber
2\Delta_{c}^{-}=\frac{1}{4}+\frac{2}{\gamma}-\frac{2}{\pi}\sqrt{1-\frac{1}{H_{c}}}-\frac{4}{\pi\gamma}\sqrt{1-\frac{H}{H_{c}}}
\\ && \nonumber
2\Delta_{s}^{+}=\frac{1}{4}-\frac{2}{\gamma}-\frac{3}{2\pi}\sqrt{1-\frac{1}{H_{c}}}+\frac{16}{\pi\gamma}\sqrt{1-\frac{H}{H_{c}}}
\\ &&
2\Delta_{s}^{-}=\frac{9}{4}-\frac{6}{\gamma}-\frac{3}{2\pi}\sqrt{1-\frac{1}{H_{c}}}+\frac{32}{\pi\gamma}\sqrt{1-\frac{H}{H_{c}}}
\end{eqnarray}
for $(D_{c},D_{s})=(1/2,-1)$. The correlation function is
\begin{equation}
G^{\perp}(x,t)\approx
\frac{A_{1}\cos((k_{F\downarrow}+k_{F\uparrow})x)}{|x+iv_{c}t|^{\theta_{c1}}|x+iv_{s}t|^{\theta_{s1}}}
+\frac{A_{2}\cos((k_{F\uparrow}-k_{F\downarrow})x)}{|x+iv_{c}t|^{\theta_{c2}}|x+iv_{s}t|^{\theta_{s2}}}
\end{equation}
where
\begin{eqnarray}
\nonumber \theta_{c1} &=&
\frac{1}{2}+\frac{8}{\pi\gamma}\sqrt{1-\frac{H}{H_{c}}} \\
\nonumber \theta_{c2} &=&
\frac{1}{2}-\frac{4}{\pi}\sqrt{1-\frac{H}{H_{c}}}+\frac{8}{\pi\gamma}\sqrt{1-\frac{H}{H_{c}}}
\\ \nonumber \theta_{s1} &=&
\frac{1}{2}+\frac{1}{\pi}\sqrt{1-\frac{H}{H_{c}}}
\\
\theta_{s2} &=&
\frac{5}{2}-\frac{8}{\gamma}-\frac{3}{\pi}\sqrt{1-\frac{H}{H_{c}}}+\frac{48}{\pi\gamma}\sqrt{1-\frac{H}{H_{c}}}.
\end{eqnarray}

(vi) $G_{p}(x,t)$: The conformal dimensions are
\begin{eqnarray}
&& \nonumber
2\Delta_{c}^{+}=\frac{9}{4}-\frac{6}{\gamma}+\frac{12}{\pi\gamma}\sqrt{1-\frac{H}{H_{c}}}
\\ && \nonumber 2\Delta_{c}^{-}=\frac{1}{4}-\frac{2}{\gamma}-\frac{4}{\pi\gamma}\sqrt{1-\frac{H}{H_{c}}} \\ &&
\nonumber
2\Delta_{s}^{+}=\frac{1}{4}+\frac{2}{\gamma}-\frac{3}{2\pi}\sqrt{1-\frac{H}{H_{c}}}-\frac{8}{\pi\gamma}\sqrt{1-\frac{H}{H_{c}}}
\\ &&
2\Delta_{s}^{-}=\frac{1}{4}-\frac{2}{\gamma}-\frac{3}{2\pi}\sqrt{1-\frac{H}{H_{c}}}+\frac{8}{\pi\gamma}\sqrt{1-\frac{H}{H_{c}}}
\end{eqnarray}
for $(D_{c},D_{s})=(1/2,0)$ and
\begin{eqnarray}
&& \nonumber
2\Delta_{c}^{+}=\frac{9}{4}-\frac{6}{\gamma}-\frac{6}{\pi}\sqrt{1-\frac{H}{H_{c}}}+\frac{20}{\pi\gamma}\sqrt{1-\frac{H}{H_{c}}}
\\ && \nonumber
2\Delta_{c}^{-}=\frac{1}{4}-\frac{2}{\gamma}+\frac{2}{\pi}\sqrt{1-\frac{H}{H_{c}}}-\frac{12}{\pi\gamma}\sqrt{1-\frac{H}{H_{c}}}
\\ && \nonumber
2\Delta_{s}^{+}=\frac{1}{4}-\frac{2}{\gamma}+\frac{1}{2\pi}\sqrt{1-\frac{H}{H_{c}}}+\frac{8}{\pi\gamma}\sqrt{1-\frac{H}{H_{c}}}
\\ &&
2\Delta_{s}^{-}=\frac{9}{4}-\frac{6}{\gamma}-\frac{15}{2\pi}\sqrt{1-\frac{H}{H_{c}}}+\frac{40}{\pi\gamma}\sqrt{1-\frac{H}{H_{c}}}
\end{eqnarray}
for $(D_{c},D_{s})=(1/2,-1)$.
The correlation function is then given by
\begin{equation}
G_{p}(x,t)\approx
\frac{A_{1}\cos((k_{F\downarrow}+k_{F\uparrow})x}{|x+iv_{c}t|^{\theta_{c1}}|x+iv_{s}t|^{\theta_{s1}}}
+\frac{A_{2}\cos((k_{F\uparrow}-k_{F\downarrow})x)}{|x+iv_{c}t|^{\theta_{c2}}|x+iv_{s}t|^{\theta_{s2}}}
\end{equation}
where
\begin{eqnarray}
\theta_{c1} &=&
\frac{5}{2}-\frac{8}{\gamma}+\frac{8}{\pi\gamma}\sqrt{1-\frac{H}{H_{c}}}
\\ \nonumber \theta_{c2} &=&
\frac{5}{2}-\frac{8}{\gamma}-\frac{4}{\pi}\sqrt{1-\frac{H}{H_{c}}}+\frac{8}{\pi\gamma}\sqrt{1-\frac{H}{H_{c}}}
\\ \nonumber
\theta_{s1} &=& \frac{1}{2}-\frac{3}{\pi}\sqrt{1-\frac{H}{H_{c}}} \\
\theta_{s2} &=&
\frac{5}{2}-\frac{8}{\gamma}-\frac{7}{\pi}\sqrt{1-\frac{H}{H_{c}}}+\frac{48}{\pi\gamma}\sqrt{1-\frac{H}{H_{c}}}.
\end{eqnarray}

Finally we note that the charge and spin velocities can be derived easily from the
relations
\begin{equation}
v_{c}=\frac{\varepsilon'(k_{0})}{2\pi\rho_{c}(k_{0})},\qquad
v_{s}=\frac{\phi_{1}'(\lambda_{0})}{2\pi\rho_{s}(\lambda_{0})}.
\end{equation}
The leading terms in the velocities are then found to be
\begin{equation}
v_{c}=2\pi
n_c\left(1-\frac{12}{\pi\gamma}\sqrt{1-\frac{H}{H_{c}}}\right)
\end{equation}
and
\begin{equation}
v_{s}=\frac{H_{c}}{n_c}\sqrt{1-\frac{H}{H_{c}}}.
\end{equation}

\end{document}